\documentclass[preprint2,tighten,times]{aastex61largeauthor}
\pdfoutput=1 

\usepackage{acronym}
\usepackage{amsfonts}
\usepackage{amsmath}
\usepackage{amssymb}
\usepackage{graphicx}
\usepackage{hyperref}
\usepackage{xcolor}

\newcommand{\kms}{\ensuremath{\mathrm{km} \, \mathrm{s}^{-1}}}
\newcommand{\Mpc}{\ensuremath{\mathrm{Mpc}}}

\newcommand{\planckHnaught}{67.74}
\newcommand{\planckHnaughtUncert}{0.46}
\newcommand{\recessionCluster}{3327} 
\newcommand{\recessionClusterUncert}{72} 

\newcommand{\riessHnaught}{73.24}
\newcommand{\riessHnaughtUncert}{1.74}
\newcommand{\tfdist}{41.1}
\newcommand{\tfdistUncert}{5.8}

\newcommand{\dgwPosteriorMedian}{41.1}
\newcommand{\dgwPosteriorNinetyUpperDiff}{5.6}
\newcommand{\dgwPosteriorNinetyLowerDiff}{12.6}
\newcommand{\dgwPosteriorOneSigmaUpperDiff}{4.0}
\newcommand{\dgwPosteriorOneSigmaLowerDiff}{7.3}
\newcommand{\dgwPosteriorMAPMax}{43.8}

\newcommand{\dgwPosteriorMAPNinetyLowerDiff}{13.1}
\newcommand{\dgwPosteriorMAPOneSigmaUpperDiff}{2.9}
\newcommand{\dgwPosteriorMAPOneSigmaLowerDiff}{6.9}
\newcommand{\Hnaught}{78}
\newcommand{\HnaughtUncert}{15}
\newcommand{\HnaughtMedian}{74.0}
\newcommand{\HnaughtNinetyUpperDiff}{33}
\newcommand{\HnaughtNinetyLowerDiff}{12}
\newcommand{\HnaughtOneSigmaUpperDiff}{16.0}
\newcommand{\HnaughtOneSigmaLowerDiff}{8.0}
\newcommand{\HnaughtMAPMax}{70.0}
\newcommand{\HnaughtMAPNinetyUpperDiff}{28}
\newcommand{\HnaughtMAPNinetyLowerDiff}{11}
\newcommand{\HnaughtMAPOneSigmaUpperDiff}{12.0}
\newcommand{\HnaughtMAPOneSigmaLowerDiff}{8.0}
\newcommand{\cosiotaMedian}{-0.88}
\newcommand{\cosiotaNinetyUpperDiff}{0.32}
\newcommand{\cosiotaNinetyLowerDiff}{0.11}
\newcommand{\cosiotaOneSigmaUpperDiff}{0.18}
\newcommand{\cosiotaOneSigmaLowerDiff}{0.09}
\newcommand{\cosiotaArcCosMedian}{152}
\newcommand{\cosiotaArcCosNinetyUpperDiff}{20}
\newcommand{\cosiotaArcCosNinetyLowerDiff}{27}
\newcommand{\cosiotaArcCosOneSigmaUpperDiff}{14}
\newcommand{\cosiotaArcCosOneSigmaLowerDiff}{17}
\newcommand{\cosiotaMAPMax}{-0.974}
\newcommand{\cosiotaMAPNinetyUpperDiff}{0.332}
\newcommand{\cosiotaMAPNinetyLowerDiff}{0.026}
\newcommand{\cosiotaMAPOneSigmaUpperDiff}{0.164}
\newcommand{\cosiotaMAPOneSigmaLowerDiff}{0.026}
\newcommand{\cosiotaMAPArcCosMax}{167}
\newcommand{\cosiotaMAPArcCosNinetyUpperDiff}{13}
\newcommand{\cosiotaMAPArcCosNinetyLowerDiff}{37}
\newcommand{\cosiotaMAPArcCosOneSigmaUpperDiff}{13}
\newcommand{\cosiotaMAPArcCosOneSigmaLowerDiff}{23}
\newcommand{\cosiotaMAPOneSigmaUpper}{-0.81}
\newcommand{\cosiotaMAPOneSigmaLower}{-1.00}
\newcommand{\cosiotaMAPArcCosOneSigmaUpper}{180}
\newcommand{\cosiotaMAPArcCosOneSigmaLower}{144}
\newcommand{\pecvel}{310}
\newcommand{\pecvelUncert}{150}
\newcommand{\pecvelstderr}{69}
\newcommand{\planckCosIotaMedian}{-0.948}
\newcommand{\planckCosIotaNinetyUpperDiff}{0.091}
\newcommand{\planckCosIotaNinetyLowerDiff}{0.046}
\newcommand{\planckCosIotaOneSigmaUpperDiff}{0.052}
\newcommand{\planckCosIotaOneSigmaLowerDiff}{0.036}
\newcommand{\planckCosIotaArcCosMedian}{161.0}
\newcommand{\planckCosIotaArcCosNinetyUpperDiff}{12}
\newcommand{\planckCosIotaArcCosNinetyLowerDiff}{12}
\newcommand{\planckCosIotaArcCosOneSigmaUpperDiff}{8.0}
\newcommand{\planckCosIotaArcCosOneSigmaLowerDiff}{8.0}
\newcommand{\riessCosIotaMedian}{-0.901}
\newcommand{\riessCosIotaNinetyUpperDiff}{0.106}
\newcommand{\riessCosIotaNinetyLowerDiff}{0.083}
\newcommand{\riessCosIotaOneSigmaUpperDiff}{0.065}
\newcommand{\riessCosIotaOneSigmaLowerDiff}{0.057}
\newcommand{\riessCosIotaArcCosMedian}{154.0}
\newcommand{\riessCosIotaArcCosNinetyUpperDiff}{15}
\newcommand{\riessCosIotaArcCosNinetyLowerDiff}{12}
\newcommand{\riessCosIotaArcCosOneSigmaUpperDiff}{9.0}
\newcommand{\riessCosIotaArcCosOneSigmaLowerDiff}{8.0}
\newcommand{\planckCosIotaMAPMax}{-0.982}
\newcommand{\planckCosIotaMAPNinetyUpperDiff}{0.104}
\newcommand{\planckCosIotaMAPNinetyLowerDiff}{0.018}
\newcommand{\planckCosIotaMAPOneSigmaUpperDiff}{0.060}
\newcommand{\planckCosIotaMAPOneSigmaLowerDiff}{0.016}
\newcommand{\planckCosIotaMAPArcCosMax}{169.0}
\newcommand{\planckCosIotaMAPArcCosNinetyUpperDiff}{11}
\newcommand{\planckCosIotaMAPArcCosNinetyLowerDiff}{18}
\newcommand{\planckCosIotaMAPArcCosOneSigmaUpperDiff}{8.0}
\newcommand{\planckCosIotaMAPArcCosOneSigmaLowerDiff}{12.0}
\newcommand{\planckCosIotaMAPOneSigmaUpper}{-0.92}
\newcommand{\planckCosIotaMAPOneSigmaLower}{-1.00}
\newcommand{\planckCosIotaMAPArcCosOneSigmaUpper}{177}
\newcommand{\planckCosIotaMAPArcCosOneSigmaLower}{157}
\newcommand{\riessCosIotaMAPMax}{-0.912}
\newcommand{\riessCosIotaMAPNinetyUpperDiff}{0.095}
\newcommand{\riessCosIotaMAPNinetyLowerDiff}{0.086}
\newcommand{\riessCosIotaMAPOneSigmaUpperDiff}{0.061}
\newcommand{\riessCosIotaMAPOneSigmaLowerDiff}{0.059}
\newcommand{\riessCosIotaMAPArcCosMax}{156.0}
\newcommand{\riessCosIotaMAPArcCosNinetyUpperDiff}{21}
\newcommand{\riessCosIotaMAPArcCosNinetyLowerDiff}{11}
\newcommand{\riessCosIotaMAPArcCosOneSigmaUpperDiff}{10.0}
\newcommand{\riessCosIotaMAPArcCosOneSigmaLowerDiff}{7.0}
\newcommand{\riessCosIotaMAPOneSigmaUpper}{-0.85}
\newcommand{\riessCosIotaMAPOneSigmaLower}{-0.97}
\newcommand{\riessCosIotaMAPArcCosOneSigmaUpper}{166}
\newcommand{\riessCosIotaMAPArcCosOneSigmaLower}{148}
\newcommand{\HubbleVelMAPMax}{3017}
\newcommand{\HubbleVelMAPOneSigmaUpperDiff}{166}

\newcommand{\HnaughtLargeVelocityMedian}{74.0}
\newcommand{\HnaughtLargeVelocityNinetyUpperDiff}{33}
\newcommand{\HnaughtLargeVelocityNinetyLowerDiff}{14}
\newcommand{\HnaughtLargeVelocityOneSigmaUpperDiff}{16.0}
\newcommand{\HnaughtLargeVelocityOneSigmaLowerDiff}{9.0}
\newcommand{\HnaughtLargeVelocityMAPMax}{70.0}
\newcommand{\HnaughtLargeVelocityMAPNinetyUpperDiff}{29}
\newcommand{\HnaughtLargeVelocityMAPNinetyLowerDiff}{14}
\newcommand{\HnaughtLargeVelocityMAPOneSigmaUpperDiff}{14.0}
\newcommand{\HnaughtLargeVelocityMAPOneSigmaLowerDiff}{9.0}
\newcommand{\HnaughtFlatZMedian}{81}
\newcommand{\HnaughtFlatZNinetyUpperDiff}{50}
\newcommand{\HnaughtFlatZNinetyLowerDiff}{17}
\newcommand{\HnaughtFlatZOneSigmaUpperDiff}{27}
\newcommand{\HnaughtFlatZOneSigmaLowerDiff}{13}
\newcommand{\HnaughtFlatZMAPMax}{71.0}
\newcommand{\HnaughtFlatZMAPNinetyUpperDiff}{48}
\newcommand{\HnaughtFlatZMAPNinetyLowerDiff}{11}
\newcommand{\HnaughtFlatZMAPOneSigmaUpperDiff}{23.0}
\newcommand{\HnaughtFlatZMAPOneSigmaLowerDiff}{9.0}

\newcommand{\hubblevelcarrick}{3047}
\newcommand{\hubblevelUncertcarrick}{150}

\newcommand{\gwdata}{\ensuremath{x_{\mathrm{GW}}}}

\begin{document}

\title{A gravitational-wave standard siren measurement of the Hubble constant}

\AuthorCallLimit=7

\author{The LIGO Scientific Collaboration and The Virgo Collaboration}
\noaffiliation
\author{The 1M2H Collaboration}
\noaffiliation
\author{The Dark Energy Camera GW-EM Collaboration and the DES Collaboration}
\noaffiliation
\author{The DLT40 Collaboration}
\noaffiliation
\author{The Las Cumbres Observatory Collaboration}
\noaffiliation
\author{The VINROUGE Collaboration}
\noaffiliation
\author{The MASTER Collaboration}
\noaffiliation

\author{B.~P.~Abbott}
\affiliation{LIGO, California Institute of Technology, Pasadena, CA 91125, USA}
\author{R.~Abbott}
\affiliation{LIGO, California Institute of Technology, Pasadena, CA 91125, USA}
\author{T.~D.~Abbott}
\affiliation{Louisiana State University, Baton Rouge, LA 70803, USA}
% ...
\collaboration{The LIGO Scientific Collaboration and The Virgo Collaboration}

\author{R.~J.~Foley}
\affiliation{Department of Astronomy and Astrophysics, University of California, Santa Cruz, CA 95064, USA}
\author{D.~A.~Coulter}
\affiliation{Department of Astronomy and Astrophysics, University of California, Santa Cruz, CA 95064, USA}
\author{M.~R.~Drout}
\affiliation{The Observatories of the Carnegie Institution for Science, 813 Santa Barbara Street, Pasadena, CA 91101}
\affiliation{Hubble and Carnegie-Dunlap Fellow}
\author{D.~Kasen}
\affiliation{Nuclear Science Division, Lawrence Berkeley National Laboratory, Berkeley, CA 94720, USA}
\affiliation{Departments of Physics and Astronomy, University of California, Berkeley, CA 94720, USA}
\author{C.~D.~Kilpatrick}
\affiliation{Department of Astronomy and Astrophysics, University of California, Santa Cruz, CA 95064, USA}
\author{B.~F.~Madore}
\affiliation{The Observatories of the Carnegie Institution for Science, 813 Santa Barbara Street, Pasadena, CA 91101}
\author{A.~Murguia-Berthier}
\affiliation{Department of Astronomy and Astrophysics, University of California, Santa Cruz, CA 95064, USA}
\author{Y.-C.~Pan}
\affiliation{Department of Astronomy and Astrophysics, University of California, Santa Cruz, CA 95064, USA}
\author{A.~L.~Piro}
\affiliation{The Observatories of the Carnegie Institution for Science, 813 Santa Barbara Street, Pasadena, CA 91101}
\author{J.~X.~Prochaska}
\affiliation{Department of Astronomy and Astrophysics, University of California, Santa Cruz, CA 95064, USA}
\author{E.~Ramirez-Ruiz}
\affiliation{Department of Astronomy and Astrophysics, University of California, Santa Cruz, CA 95064, USA}
\affiliation{Dark Cosmology Centre, Niels Bohr Institute, University of Copenhagen, Blegdamsvej 17, 2100 Copenhagen, Denmark}
\author{A.~Rest}
\affiliation{Space Telescope Science Institute, 3700 San Martin Drive, Baltimore, MD 21218}
\author{C.~Rojas-Bravo}
\affiliation{Department of Astronomy and Astrophysics, University of California, Santa Cruz, CA 95064, USA}
\author{B.~J.~Shappee}
\affiliation{The Observatories of the Carnegie Institution for Science, 813 Santa Barbara Street, Pasadena, CA 91101}
\affiliation{Institute for Astronomy, University of Hawai'i, 2680 Woodlawn Drive, Honolulu, HI 96822, USA}
\affiliation{Hubble and Carnegie-Princeton Fellow}
\author{M.~R.~Siebert}
\affiliation{Department of Astronomy and Astrophysics, University of California, Santa Cruz, CA 95064, USA}
\author{J.~D.~Simon}
\affiliation{The Observatories of the Carnegie Institution for Science, 813 Santa Barbara Street, Pasadena, CA 91101}
\author{N.~Ulloa}
\affiliation{Departamento de F\'{i}sica y Astronom\'{i}a, Universidad de La Serena, La Serena, Chile}

\collaboration{The 1M2H Collaboration}

\author{J.~Annis}
\affiliation{Fermi National Accelerator Laboratory, P. O. Box 500, Batavia, IL 60510, USA}
\author{M.~Soares-Santos}
\affiliation{Department of Physics, Brandeis University, Waltham MA, USA}
\affiliation{Fermi National Accelerator Laboratory, P. O. Box 500, Batavia, IL 60510, USA}
\author{D.~Brout}
\affiliation{Department of Physics and Astronomy, University of Pennsylvania, Philadelphia, PA 19104, USA}
\author{D.~Scolnic}
\affiliation{Kavli Institute for Cosmological Physics, University of Chicago, Chicago, IL 60637, USA}
\author{H.~T.~Diehl}
\affiliation{Fermi National Accelerator Laboratory, P. O. Box 500, Batavia, IL 60510, USA}
\affiliation{Fermi National Accelerator Laboratory, P. O. Box 500, Batavia, IL 60510, USA}
\author{J.~Frieman}
\affiliation{Fermi National Accelerator Laboratory, P. O. Box 500, Batavia, IL 60510, USA}
\affiliation{Kavli Institute for Cosmological Physics, University of Chicago, Chicago, IL 60637, USA}
\author{E.~Berger}
\affiliation{Harvard-Smithsonian Center for Astrophysics, 60 Garden Street, Cambridge, Massachusetts 02138, USA}
\author{K.~D.~Alexander}
\affiliation{Harvard-Smithsonian Center for Astrophysics, 60 Garden Street, Cambridge, Massachusetts 02138, USA}
\author{S.~Allam}
\affiliation{Fermi National Accelerator Laboratory, P. O. Box 500, Batavia, IL 60510, USA}
\affiliation{Fermi National Accelerator Laboratory, P. O. Box 500, Batavia, IL 60510, USA}
\author{E.~Balbinot}
\affiliation{Department of Physics, University of Surrey, Guildford, GU2 7XH, UK}
% ...

\collaboration{The Dark Energy Camera GW-EM Collaboration
  and the DES Collaboration}

\author{J.~B.~Haislip}
\affiliation{Department of Physics and Astronomy, University of North Carolina at Chapel Hill, Chapel Hill, NC, 27599, USA}
\author{V.~V.~Kouprianov}
\affiliation{Department of Physics and Astronomy, University of North Carolina at Chapel Hill, Chapel Hill, NC, 27599, USA}
\author{D.~E.~Reichart}
\affiliation{Department of Physics and Astronomy, University of North Carolina at Chapel Hill, Chapel Hill, NC, 27599, USA}
\author{L.~Tartaglia}
\affiliation{Department of Astronomy and Steward Observatory, University of Arizona, 933 N Cherry Ave, Tucson, AZ 85719, USA}
\affiliation{Department of Physics, University of California, 1 Shields Avenue, Davis, CA 95616-5270, USA}
\author{D.~J.~Sand}
\affiliation{Department of Astronomy and Steward Observatory, University of Arizona, 933 N Cherry Ave, Tucson, AZ 85719, USA}
\author{S.~Valenti}
\affiliation{Department of Physics, University of California, 1 Shields Avenue, Davis, CA 95616-5270, USA}
\author{S.~Yang}
\affiliation{Department of Physics, University of California, 1 Shields Avenue, Davis, CA 95616-5270, USA}
\affiliation{Department of Physics and Astronomy, University of Padova, Via 8 Febbraio, 2-35122 Padova, Italy}
\affiliation{INAF Osservatorio Astronomico di Padova, Vicolo della Osservatorio 5, I-35122 Padova, Italy}

\collaboration{The DLT40 Collaboration}

\author{Iair Arcavi}
\affiliation{Department of Physics, University of California, Santa Barbara, CA 93106-9530, USA}
\affiliation{Las Cumbres Observatory, 6740 Cortona Dr Ste 102, Goleta, CA 93117-5575, USA}
\author{Griffin Hosseinzadeh}
\affiliation{Department of Physics, University of California, Santa Barbara, CA 93106-9530, USA}
\affiliation{Las Cumbres Observatory, 6740 Cortona Dr Ste 102, Goleta, CA 93117-5575, USA}
\author{D. Andrew Howell}
\affiliation{Department of Physics, University of California, Santa Barbara, CA 93106-9530, USA}
\affiliation{Las Cumbres Observatory, 6740 Cortona Dr Ste 102, Goleta, CA 93117-5575, USA}
\author{Curtis McCully}
\affiliation{Department of Physics, University of California, Santa Barbara, CA 93106-9530, USA}
\affiliation{Las Cumbres Observatory, 6740 Cortona Dr Ste 102, Goleta, CA 93117-5575, USA}
\author{Dovi Poznanski}
\affiliation{School of Physics and Astronomy, Tel Aviv University, Tel Aviv 69978, Israel}
\author{Sergiy Vasylyev}
\affiliation{Department of Physics, University of California, Santa Barbara, CA 93106-9530, USA}
\affiliation{Las Cumbres Observatory, 6740 Cortona Dr Ste 102, Goleta, CA 93117-5575, USA}

\collaboration{The Las Cumbres Observatory Collaboration}

\author{N. R. Tanvir}
\affiliation{Department of Physics and Astronomy, University of Leicester, University Road, Leicester, LE1 7RH, UK}
\author{A. J. Levan}
\affiliation{Department of Physics, University of Warwick, Coventry, CV4 7AL, UK}
\author{J. Hjorth}
\affiliation{DARK, Niels Bohr Institute, University of Copenhagen, Juliane Maries Vej 30, 2100 Copenhagen \O,  Denmark}
\author{Z. Cano}
\affiliation{Instituto de Astrof\'isica de Andaluc\'ia (IAA-CSIC), Glorieta de la Astronom\'ia s/n, 18008 Granada, Spain}
\author{C. Copperwheat}
\affiliation{Astrophysics Research Institute, Liverpool John Moores University, IC2, Liverpool Science Park, 146 Brownlow Hill, Liverpool L3 5RF}
\author{A. de Ugarte-Postigo}
\affiliation{Instituto de Astrof\'isica de Andaluc\'ia (IAA-CSIC), Glorieta de la Astronom\'ia s/n, 18008 Granada, Spain}
\author{P.A. Evans}
\affiliation{Department of Physics and Astronomy, University of Leicester, University Road, Leicester, LE1 7RH, UK}
\author{J.P.U. Fynbo}
\affiliation{DARK, Niels Bohr Institute, University of Copenhagen, Juliane Maries Vej 30, 2100 Copenhagen \O,  Denmark}
\author{C. Gonz\'alez-Fern\'andez}
\affiliation{Institute of Astronomy, University of Cambridge, Madingley Road, Cambridge, CB3 0HA, United Kingdom}
\author{J. Greiner}
\affiliation{Max-Planck-Institut f\"ur extraterrestrische Physik, 85740 Garching, Giessenbachstr. 1, Germany}
\author{M. Irwin}
\affiliation{Institute of Astronomy, University of Cambridge, Madingley Road, Cambridge, CB3 0HA, United Kingdom}
\author{J. Lyman}
\affiliation{Department of Physics, University of Warwick, Coventry, CV4 7AL, UK}
\author{I. Mandel}
\affiliation{Birmingham Institute for Gravitational Wave Astronomy and School of Physics and Astronomy, University of Birmingham, Birmingham, B15 2TT, UK}
\author{R. McMahon}
\affiliation{Institute of Astronomy, University of Cambridge, Madingley Road, Cambridge, CB3 0HA, United Kingdom}
\author{B. Milvang-Jensen}
\affiliation{DARK, Niels Bohr Institute, University of Copenhagen, Juliane Maries Vej 30, 2100 Copenhagen \O,  Denmark}
\author{P. O'Brien}
\affiliation{Department of Physics and Astronomy, University of Leicester, University Road, Leicester, LE1 7RH, UK}
\author{J. P. Osborne}
\affiliation{Department of Physics and Astronomy, University of Leicester, University Road, Leicester, LE1 7RH, UK}
\author{D. A. Perley}
\affiliation{Astrophysics Research Institute, Liverpool John Moores University, IC2, Liverpool Science Park, 146 Brownlow Hill, Liverpool L3 5RF}
\author{E. Pian}
\affiliation{INAF, Institute of Space Astrophysics and Cosmic Physics, Via Gobetti 101, I-40129 Bologna, Italy}
\author{E. Palazzi}
\affiliation{INAF, Institute of Space Astrophysics and Cosmic Physics, Via Gobetti 101, I-40129 Bologna, Italy}
\author{E. Rol}
\affiliation{School of Physics and Astronomy, Monash University, VIC 3800, Australia; Monash Centre for Astrophysics, Monash University, VIC 3800, Australia}
\author{S. Rosetti}
\affiliation{Department of Physics and Astronomy, University of Leicester, University Road, Leicester, LE1 7RH, UK}
\author{S. Rosswog}
\affiliation{The Oskar Klein Centre, Department of Astronomy, AlbaNova, Stockholm University, SE-106 91 Stockholm, Sweden}
\author{A. Rowlinson}
\affiliation{Anton Pannekoek Institute, University of Amsterdam, Science Park 904, 1098 XH Amsterdam, the Netherlands}
\affiliation{ASTRON, the Netherlands Institute for Radio Astronomy, Postbus 2, 7990 AA Dwingeloo, the Netherlands}
\author{S. Schulze}
\affiliation{Department of Particle Physics and Astrophysics, Weizmann Institute of Science, 76100, Rehovot, Israel}
\author{D.T.H. Steeghs}
\affiliation{Department of Physics, University of Warwick, Coventry, CV4 7AL, UK}
\author{C.C. Th\"one}
\affiliation{Instituto de Astrof\'isica de Andaluc\'ia (IAA-CSIC), Glorieta de la Astronom\'ia s/n, 18008 Granada, Spain}
\author{K. Ulaczyk}
\affiliation{Department of Physics, University of Warwick, Coventry, CV4 7AL, UK}
\author{D. Watson}
\affiliation{DARK, Niels Bohr Institute, University of Copenhagen, Juliane Maries Vej 30, 2100 Copenhagen \O,  Denmark}
\author{K. Wiersema}
\affiliation{Department of Physics and Astronomy, University of Leicester, University Road, Leicester, LE1 7RH, UK}
\affiliation{Department of Physics, University of Warwick, Coventry, CV4 7AL, UK}

\collaboration{The VINROUGE Collaboration}

\author{V.M. Lipunov}
\affiliation{M.V.Lomonosov Moscow State University, Physics Department, Leninskie gory, GSP-1, Moscow, 119991, Russia}
\affiliation{M.V.Lomonosov Moscow State University, Sternberg Astronomical Institute, Universitetsky pr., 13, Moscow, 119234, Russia}
\author{E. Gorbovskoy}
\affiliation{M.V.Lomonosov Moscow State University, Sternberg Astronomical Institute, Universitetsky pr., 13, Moscow, 119234, Russia}
\author{V.G. Kornilov}
\affiliation{M.V.Lomonosov Moscow State University, Physics Department, Leninskie gory, GSP-1, Moscow, 119991, Russia}
\affiliation{M.V.Lomonosov Moscow State University, Sternberg Astronomical Institute, Universitetsky pr., 13, Moscow, 119234, Russia}
\author{N .Tyurina}
\affiliation{M.V.Lomonosov Moscow State University, Sternberg Astronomical Institute, Universitetsky pr., 13, Moscow, 119234, Russia}
\author{P. Balanutsa}
\affiliation{M.V.Lomonosov Moscow State University, Sternberg Astronomical Institute, Universitetsky pr., 13, Moscow, 119234, Russia}
\author{D.Vlasenko}
\affiliation{M.V.Lomonosov Moscow State University, Physics Department, Leninskie gory, GSP-1, Moscow, 119991, Russia}
\affiliation{M.V.Lomonosov Moscow State University, Sternberg Astronomical Institute, Universitetsky pr., 13, Moscow, 119234, Russia}
\author{I.Gorbunov}
\affiliation{M.V.Lomonosov Moscow State University, Sternberg Astronomical Institute, Universitetsky pr., 13, Moscow, 119234, Russia}
\author{R. Podesta}
\affiliation{Observatorio Astronomico Felix Aguilar (OAFA) , National University of San Juan, Argentina}
\author{H. Levato}
\affiliation{Instituto de Ciencias Astronomicas,de la Tierra y del Espacio (ICATE), San Juan, Argentina}
\author{C. Saffe}
\affiliation{Instituto de Ciencias Astronomicas,de la Tierra y del Espacio (ICATE), San Juan, Argentina}
\author{D.A.H.Buckley}
\affiliation{South African Astrophysical Observatory, PO Box 9, 7935 Observatory, Cape Town, South Africa}
\author{N.M. Budnev}
\affiliation{Irkutsk State University, Applied Physics Institute, 20, Gagarin blvd,664003, Irkutsk, Russia}
\author{O. Gress}
\affiliation{Irkutsk State University, Applied Physics Institute, 20, Gagarin blvd,664003, Irkutsk, Russia}
\affiliation{M.V.Lomonosov Moscow State University, Sternberg Astronomical Institute, Universitetsky pr., 13, Moscow, 119234, Russia}
\author{V. Yurkov}
\affiliation{Blagoveschensk State Pedagogical University, Lenin str., 104, Amur Region, Blagoveschensk 675000}
\author{R. Rebolo}
\affiliation{Instituto de Astrof\'isica de Canarias Via Lactea, s/n E38205 - La Laguna (Tenerife), Spain}
\author{M. Serra-Ricart}
\affiliation{Instituto de Astrof\'isica de Canarias Via Lactea, s/n E38205 - La Laguna (Tenerife), Spain}

\collaboration{The MASTER Collaboration}

\acrodef{BNS}[BNS]{binary neutron star}
\acrodef{CBC}[CBC]{compact binary coalescence}
\acrodef{GW}[GW]{gravitational-wave}
\acrodef{EOS}[EOS]{equation of state}
\acrodef{NS}[NS]{neutron star}
\acrodef{EM}[EM]{electromagnetic}
\acrodef{SNR}[SNR]{signal-to-noise-ratio}
\acrodef{ET}[ET]{Einstein Telescope}
\acrodef{GRB}[GRB]{Gamma Ray Burst}
\acrodef{SN}[SNe]{Supernovae Type Ia}
\acrodef{GBM}[GBM]{Gamma-Ray Burst Monitor}
\acrodef{CMB}[CMB]{Cosmic Microwave Background}
\acrodef{PN}[PN]{post-Newtonian}
\acrodef{LIGO}[LIGO]{Laser Interferometer Gravitational-wave
  Observatory}
\acrodef{LISA}[LISA]{laser interferometer space antenna}
\acrodef{ISCO}[ISCO]{innermost-stable-circular orbit}
\acrodef{BHNS}[BHNS]{black-hole---neutron star}
\acrodef{NR}[NR]{numerical relativity}
\acrodef{PE}[PE]{parameter estimation}
\acrodef{2MASS}[2MASS]{two micron all sky survey}
\acrodef{MAP}{maximum a posteriori}

\begin{abstract}
  The detection of GW170817 \citep{GW170817} in both gravitational
  waves and electromagnetic waves heralds the age of
  gravitational-wave multi-messenger astronomy.  On 17 August 2017 the
  Advanced \ac{LIGO} \citep{abbott2015aligo} and Virgo
  \citep{acernese2015advirgo} detectors observed GW170817, a strong
  signal from the merger of a binary neutron-star system.  Less than 2
  seconds after the merger, a gamma-ray burst (GRB 170817A) was
  detected within a region of the sky consistent with the
  LIGO-Virgo-derived location of the gravitational-wave source
  \citep{grb1,grb2,grb3}.  This sky region was subsequently observed
  by optical astronomy facilities \citep{GW170817MMA}, resulting in
  the identification of an optical transient signal within $\sim 10$
  arcsec of the galaxy NGC\,4993
  \citep{1M2H,DECam,DLT40,LCO,VISTA,MASTER}.  These multi-messenger
  observations allow us to use GW170817 as a standard siren
  \citep{Schutz1986,Holz2005,2006PhRvD..74f3006D,Nissanke2010,2013arXiv1307.2638N},
  the gravitational-wave analog of an astronomical standard candle, to
  measure the Hubble constant.  This quantity, which represents the
  local expansion rate of the Universe, sets the overall scale of the
  Universe and is of fundamental importance to cosmology.  Our
  measurement combines the distance to the source inferred purely from
  the gravitational-wave signal with the recession velocity inferred
  from measurements of the redshift using electromagnetic data.  This
  approach does not require any form of cosmic ``distance ladder''
  \citep{Freedman2001}; the \ac{GW} analysis can be used to estimate
  the luminosity distance out to cosmological scales directly, without
  the use of intermediate astronomical distance measurements. We
  determine the Hubble constant to be
  $\HnaughtMAPMax^{+\HnaughtMAPOneSigmaUpperDiff}_{-\HnaughtMAPOneSigmaLowerDiff}
  \, \kms \, \Mpc^{-1}$ (maximum a posteriori and 68\% credible
  interval).  This is consistent with existing measurements
  \citep{Planck2016,Riess2016}, while being completely independent of
  them.  Additional standard-siren measurements from future
  gravitational-wave sources will provide precision constraints of
  this important cosmological parameter.
\end{abstract}

\section*{}

The Hubble constant $H_0$ measures the mean expansion rate of the
Universe. At nearby distances ($d\lesssim 50\,\mbox{Mpc}$) it is well
approximated by the expression
\begin{equation}
v_H=H_0d,
\label{eq:hubble}
\end{equation}
where $v_H$ is the local ``Hubble flow'' velocity of a source, and $d$
is the distance to the source.  At such distances all cosmological
distance measures (such as luminosity distance and comoving distance)
differ at the order of $v_{H}/c$ where $c$ is the speed of light. As
$v_{H}/c \sim 1\%$ for GW170817 we do not distinguish between them. We
are similarly insensitive to the values of other cosmological
parameters, such as $\Omega_m$ and $\Omega_\Lambda$.

To obtain the Hubble flow velocity at the position of GW170817, we use
the optical identification of the host galaxy
NGC\,4993 \citep{GW170817MMA}. This identification is based solely on
the 2-dimensional projected offset and is independent of any assumed
value of $H_0$.  The position and redshift of this galaxy allow us to
estimate the appropriate value of the Hubble flow velocity.  Because
the source is relatively nearby the random relative motions of
galaxies, known as peculiar velocities, need to be taken into
account. The peculiar velocity is $\sim10\%$ of the measured
recessional velocity (see Methods).

The original standard siren proposal \citep{Schutz1986} did not rely on
the unique identification of a host galaxy. By combining information
from $\sim100$ independent \ac{GW} detections, each with a set of potential host
galaxies, a $\sim 5\%$ estimate of $H_0$ can be obtained even without the detection of any transient
optical counterparts \citep{DelPozzo:2011yh}.
This is particularly relevant, as gravitational-wave networks will detect many binary black hole mergers over the coming years \citep{O1-BBH}, and these are not expected to be accompanied by electromagnetic counterparts.
Alternatively, if an EM counterpart has been identified but the host galaxy is unknown, the same statistical method can be applied but using only those
galaxies in a narrow beam around the location of the optical counterpart.
However, such statistical analyses are sensitive to a number of complicating
effects, including the incompleteness of current galaxy
catalogs or the need for dedicated follow-up surveys, as
well as a range of selection effects \citep{Messenger2013}. In what follows we
exploit the identification of NGC\,4993 as the host galaxy of GW170817 to
perform a standard siren measurement of the Hubble constant \citep{Holz2005,2006PhRvD..74f3006D,Nissanke2010,2013arXiv1307.2638N}.

Analysis of the \ac{GW} data associated with GW170817 produces
estimates for the parameters of the source, under the assumption that
general relativity is the correct model of gravity \citep{GW170817}.  We
are most interested in the joint posterior distribution on the
luminosity distance and binary orbital inclination angle. For the
analysis in this paper we fix the location of the \ac{GW} source on
the sky to the identified location of the counterpart \citep{1M2H}.
See the Methods section for details.

An analysis of the GW data alone finds
that GW170817 occurred at a distance
$d=\dgwPosteriorMAPMax^{+\dgwPosteriorMAPOneSigmaUpperDiff}_{-\dgwPosteriorMAPOneSigmaLowerDiff}$\,Mpc (all values are quoted as the maximum posterior value with the minimal width $68.3\%$ credible interval). We note that the
distance quoted here differs from that in other studies \citep{GW170817}, since
here we assume that the optical counterpart represents the true sky
location of the \ac{GW} source instead of marginalizing over a range of potential sky locations.
The $\sim 15\%$ uncertainty is due to a combination of
statistical measurement error from the noise in the detectors,
instrumental calibration uncertainties \citep{GW170817}, and a geometrical factor
dependent upon the correlation of distance with inclination angle. The GW
measurement is consistent with the distance to NGC\,4993 measured
using the Tully-Fisher relation,
$d_{\mathrm{TF}}=\tfdist \pm \tfdistUncert
\,\mathrm{Mpc}$ \citep{Sakai2000,Freedman2001}.

The measurement of the \ac{GW} polarization is crucial for inferring the binary
inclination. This inclination, $\iota$, is defined as the angle between the
line of sight vector from the source to the detector and the orbital angular momentum
vector of the binary system.  For \ac{EM} phenomena it is
typically not possible to tell whether a system is orbiting clockwise or counter-clockwise (or, equivalently, face-on or face-off), and sources are therefore
usually characterized by a viewing angle: $\min\left( \iota , 180^\circ - \iota
\right)$. By contrast, \ac{GW} measurements can identify the sense of the rotation, and thus $\iota$ ranges from $0$ (counter-clockwise) to $180\deg$ (clockwise).
Previous \ac{GW} detections by \ac{LIGO} 
had large uncertainties in luminosity distance and
inclination \citep{O1-BBH} because the two LIGO
detectors that were involved are nearly co-aligned, preventing a
precise polarization measurement. In the present case, thanks to Virgo
as an additional detector, the cosine of the inclination can be
constrained at $68.3\%$ ($1\sigma$) confidence to the range
$\left[\cosiotaMAPOneSigmaLower, \cosiotaMAPOneSigmaUpper\right]$
corresponding to inclination angles between
$\left[\cosiotaMAPArcCosOneSigmaLower,\cosiotaMAPArcCosOneSigmaUpper\right]\deg$.
This implies that the plane of the binary orbit is almost, but not
quite, perpendicular to our line of sight to the source
($\iota\approx180\deg$), which is consistent with the observation of a
coincident GRB (LVC, GBM, \& INTEGRAL 2017 in prep.; Goldstein et al. 2017, ApJL, submitted;
Savchenko et al. 2017, ApJL, submitted).  We report inferences on
$\cos \iota$ because our prior for it is flat, so the posterior is
proportional to the marginal likelihood for it from the \ac{GW}
observations.

EM follow-up of the GW sky localization region \citep{GW170817MMA} discovered an optical
transient \citep{1M2H,DECam,DLT40,LCO,VISTA,MASTER} in close
proximity to the galaxy NGC\,4993.
The location of the transient was previously
observed by the {\it Distance Less Than 40 Mpc} (DLT40) survey on 2017
July 27.99 UT and no sources were found \citep{DLT40}.
We estimate the probability of a random chance association between the optical
counterpart and NGC\,4993 to be $0.004\%$ (see the Methods section for
details). In what follows we assume that the optical counterpart is associated with GW170817, and that this source resides in NGC\,4993.

\begin{figure}
\begin{centering}
\includegraphics[width=\columnwidth]{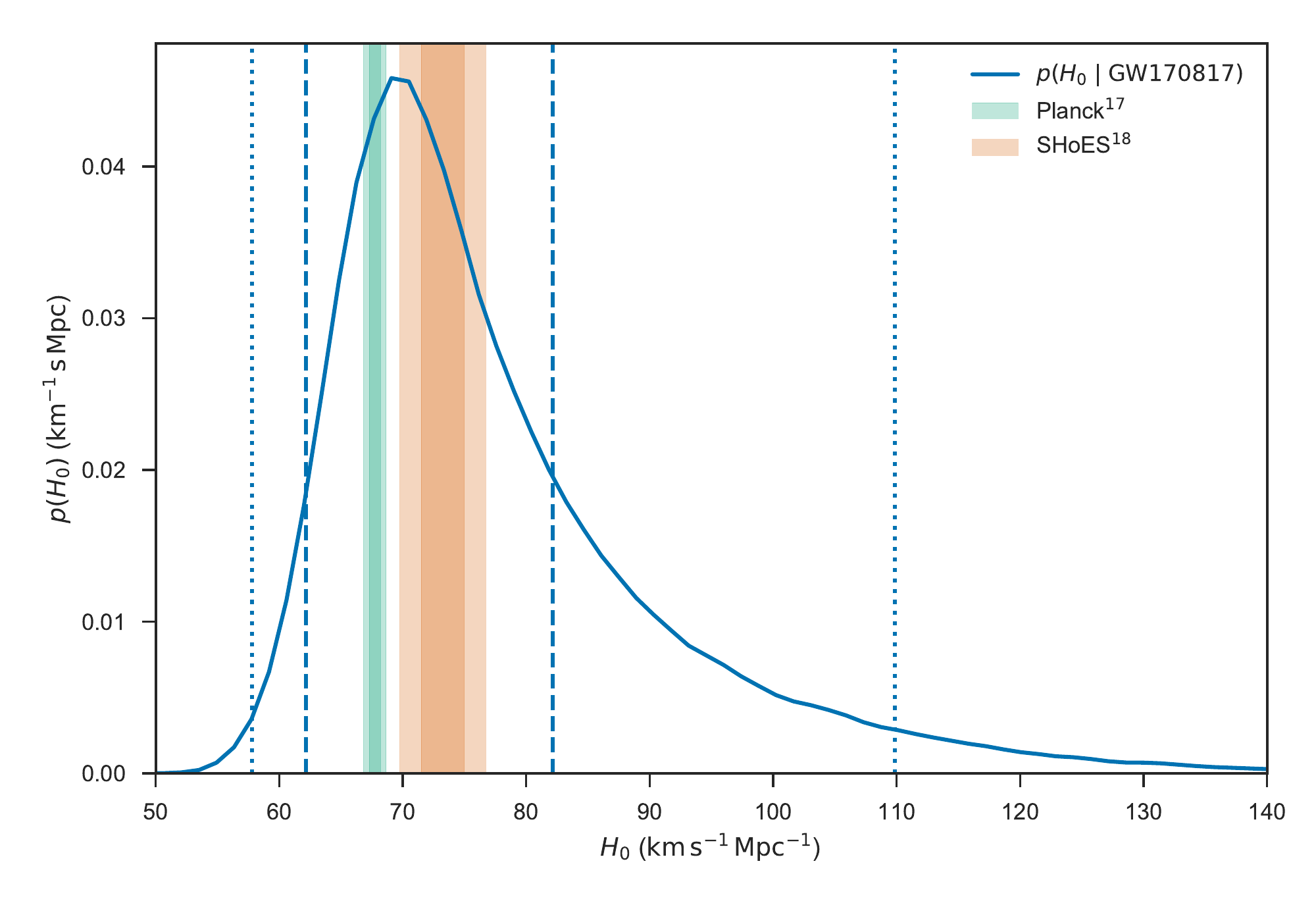}
\end{centering}
\caption{\label{fig:killer-plot1} \textbf{GW170817 measurement of
    $H_0$.}  Marginalized posterior density for $H_0$ (blue curve).
  Constraints at $1$- and $2\sigma$ from Planck \citep{Planck2016} and
  SHoES \citep{Riess2016} are shown in green and orange.  The maximum a
  posteriori value and minimal $68.3\%$ credible interval from this
  PDF is
  $H_0 =
  \HnaughtMAPMax^{+\HnaughtMAPOneSigmaUpperDiff}_{-\HnaughtMAPOneSigmaLowerDiff}\,
  \kms \, \Mpc^{-1}$.  The $68.3\%$ ($1\sigma$) and $95.4\%$
  ($2\sigma$) minimal credible intervals are indicated by dashed and
  dotted lines.}
\end{figure}

To compute $H_0$ we need to estimate the background Hubble flow velocity at the
position of NGC\,4993.  In the traditional electromagnetic calibration of the
cosmic ``distance ladder" \citep{Freedman2001}, this step is commonly carried
out using secondary distance indicator information, such as the Tully-Fisher
relation \citep{Sakai2000}, which allows one to infer the background Hubble flow
velocity in the local Universe scaled back from more distant secondary
indicators calibrated in quiet Hubble flow.  We do not adopt this approach
here, however, in order to preserve more fully the independence of our results
from the electromagnetic distance ladder.  Instead we estimate the Hubble flow
velocity at the position of NGC\,4993 by correcting for local peculiar motions.

\begin{figure}
\begin{centering}
\includegraphics[width=\columnwidth]{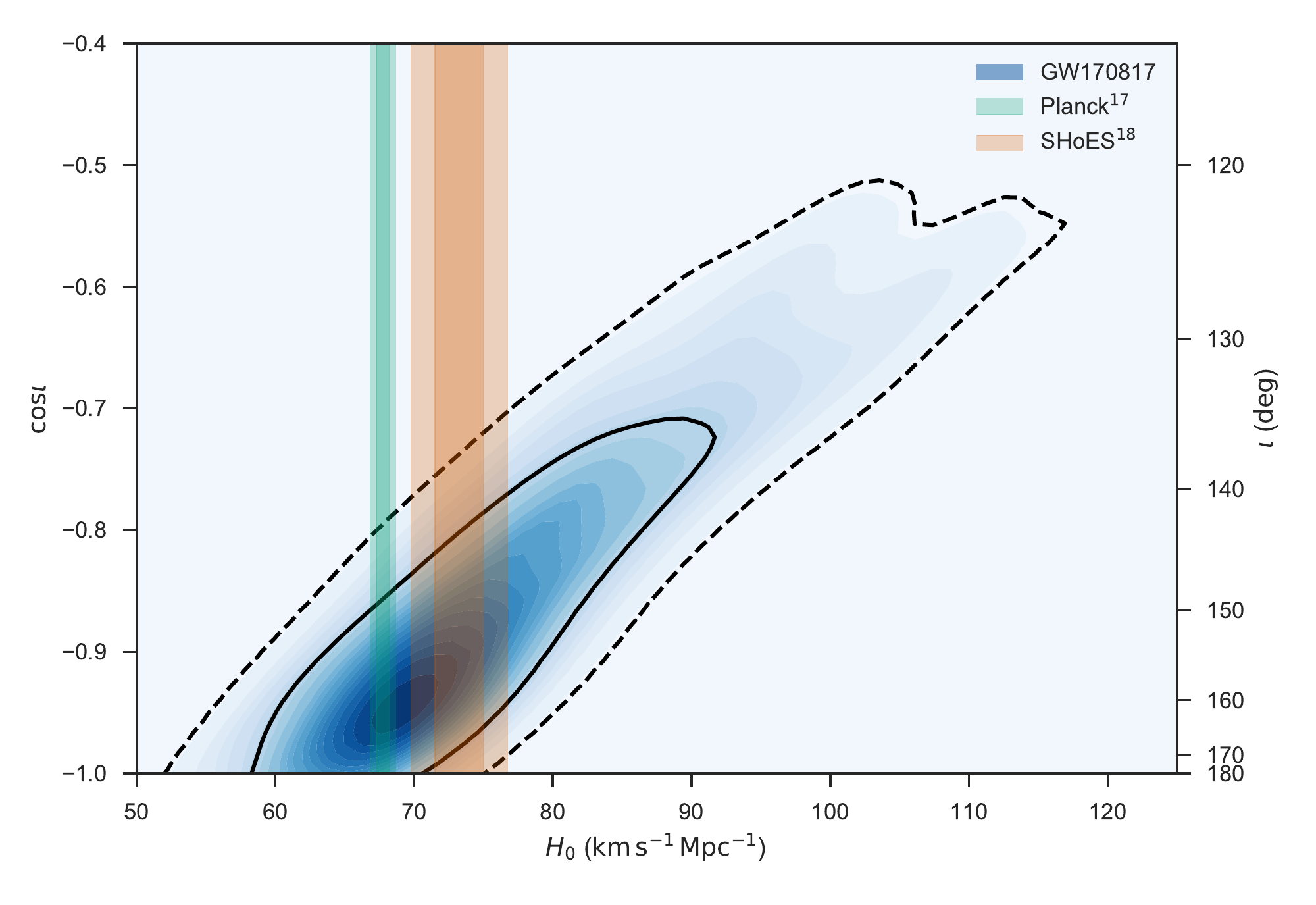}
\end{centering}
\caption{\label{fig:killer-plot2} \textbf{Inference on $H_0$ and
    inclination.} Posterior density of $H_0$ and $\cos \iota$ from the
  joint GW-EM analysis (blue contours).  Shading levels are drawn at
  every $5\%$ credible level, with the $68.3\%$ ($1\sigma$, solid) and
  $95.4\%$ ($2\sigma$, dashed) contours in black. Values of $H_0$ and
  $1$- and $2\sigma$ error bands are also displayed from
  Planck \citep{Planck2016} and SHoES \citep{Riess2016}. As noted in the
  text, inclination angles near $180\deg$ ($\cos \iota = -1$) indicate
  that the orbital angular momentum is anti-parallel with the
  direction from the source to the detector.}
\end{figure}

NGC\,4993 is part of a collection of galaxies, ESO-508, whose
center-of-mass recession velocity relative to the frame of the CMB \citep{Hinshaw2009} is \citep{Crook2007}
$\recessionCluster \pm \recessionClusterUncert \, \kms$.  We correct
the group velocity by $\pecvel \, \kms$ due to the coherent bulk
flow \citep{Springob2014,Carrick2015} towards The Great Attractor (see
Methods section for details).  The standard error on our estimate of
the peculiar velocity is $\pecvelstderr \, \kms$, but recognizing that
this value may be sensitive to details of the bulk flow motion that
have been imperfectly modelled, in our subsequent analysis we adopt a
more conservative estimate \citep{Carrick2015} of $150 \kms$ for the
uncertainty on the peculiar velocity at the location of NGC\,4993, and
fold this into our estimate of the uncertainty on $v_H$.
From this, we obtain a Hubble velocity $v_H =
\HubbleVelMAPMax\pm\HubbleVelMAPOneSigmaUpperDiff\,\kms$.

Once the distance and Hubble velocity distributions have been determined from
the \ac{GW} and \ac{EM} data, respectively, we can constrain the value of the
Hubble constant.  The measurement of the distance is strongly correlated with
the measurement of the inclination of the orbital plane of the binary.  The
analysis of the \ac{GW} data also depends on other parameters describing the
source, such as the masses of the components \citep{O1-BBH}. Here we treat the uncertainty in these other variables
by marginalizing over the posterior distribution on system
parameters \citep{GW170817}, with the exception of the position of the system on
the sky which is taken to be fixed at the location of the optical counterpart.

We carry out a Bayesian analysis to infer a posterior distribution on $H_0$ and
inclination, marginalized over uncertainties in the recessional and peculiar
velocities; see the Methods section for details.
Figure~\ref{fig:killer-plot1} shows the marginal
posterior for $H_0$.
The maximum a posteriori value with the minimal $68.3\%$ credible interval is
$H_0 = \HnaughtMAPMax^{+\HnaughtMAPOneSigmaUpperDiff}_{-\HnaughtMAPOneSigmaLowerDiff}\, \kms \, \Mpc^{-1}$.
Our estimate agrees well with state-of-the-art determinations of this quantity,
including CMB measurements from Planck \citep{Planck2016} ($\planckHnaught \pm
\planckHnaughtUncert \, \kms \, \Mpc^{-1}$, ``TT,TE,EE+lowP+lensing+ext'') and Type Ia supernova measurements from SHoES \citep{Riess2016}
($\riessHnaught \pm \riessHnaughtUncert \, \kms \, \Mpc^{-1}$), as well as baryon acoustic oscillations measurements from SDSS \citep{2015PhRvD..92l3516A}, strong lensing measurements from H0LiCOW \citep{2017MNRAS.465.4914B}, high-$l$ CMB measurements from SPT \citep{2017arXiv170709353H}, and Cepheid measurements from the HST key project \citep{Freedman2001}.  Our measurement is
a new and independent determination of this quantity.
The close agreement indicates that, although each method may be affected by
different systematic uncertainties, we see no evidence at present for a
systematic difference between \ac{GW} and established \ac{EM}-based estimates.  As
has been much remarked upon, the Planck and SHoES results are inconsistent at
$\gtrsim3\sigma$ level. Our measurement does not resolve this tension,
and is broadly consistent with both.

One of the main sources of uncertainty in our measurement of $H_0$ is
due to the degeneracy between distance and inclination in the \ac{GW}
measurements. A face-on or face-off binary far away has a similar
gravitational-wave amplitude to an edge-on binary closer in. This relationship is
captured in Figure~\ref{fig:killer-plot2}, which shows posterior
contours in the $H_0$--$\cos\iota$ parameter space.

\begin{figure}
\begin{centering}
\includegraphics[width=\columnwidth]{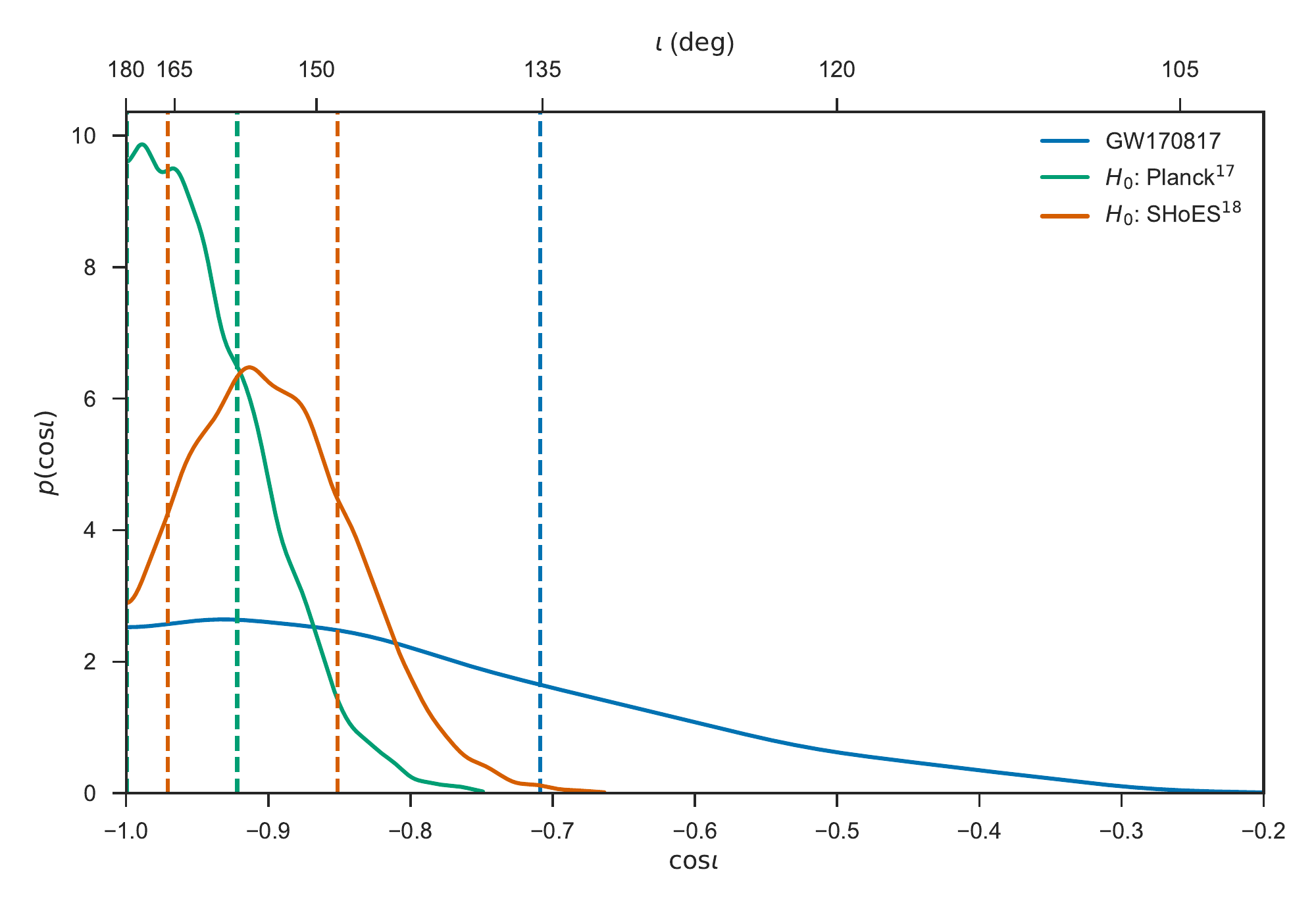}
\end{centering}
\caption{\label{fig:inclination} \textbf{Constraints on the
    inclination angle of GW170817.}  Posterior density on
  $\cos \iota$, for various assumptions about the prior distribution
  of $H_0$.  The analysis of the joint \ac{GW} and \ac{EM} data with a
  $1/H_0$ prior density gives the blue curve; using values of $H_0$
  from Planck \citep{Planck2016} and SHoES \citep{Riess2016} as a prior on
  $H_0$ give the green and red curves, respectively.  Choosing a
  narrow prior on $H_0$ converts the precise Hubble velocity
  measurements for the group containing NGC\,4993 to a precise
  distance measurement, breaking the distance inclination degeneracy,
  and leading to strong constraints on the inclination.  Minimal
  $68.3\%$ ($1\sigma$) credible intervals are indicated by dashed
  lines.  Because our prior on inclination is flat on $\cos \iota$ the
  densities in this plot are proportional to the marginalised
  likelihood for $\cos \iota$.  }
\end{figure}

The posterior in Figure~\ref{fig:killer-plot1} results from the
vertical projection of Figure~\ref{fig:killer-plot2}, marginalizing
out uncertainties in the cosine of inclination to derive constraints on the
Hubble constant. Alternatively, it is possible to project
horizontally, and thereby marginalize out the Hubble constant to
derive constraints on the cosine of inclination.  If instead of deriving $H_0$
independently we take the existing constraints on
$H_0$ \citep{Planck2016,Riess2016} as priors, we are able to
significantly improve our constraints on $\cos\iota$ as shown in Figure~\ref{fig:inclination}.
Assuming the Planck value for $H_0$, the minimal 68.3\% credible
interval for the cosine of inclination is
$\left[\planckCosIotaMAPOneSigmaLower,
  \planckCosIotaMAPOneSigmaUpper\right]$ (corresponding to an
inclination angle range
$\left[\planckCosIotaMAPArcCosOneSigmaLower,\planckCosIotaMAPArcCosOneSigmaUpper\right]\deg$).
For the SHoES value of $H_0$, it is
$\left[\riessCosIotaMAPOneSigmaLower,
  \riessCosIotaMAPOneSigmaUpper\right]$ (corresponding to an
inclination angle range
$\left[\riessCosIotaMAPArcCosOneSigmaLower,\riessCosIotaMAPArcCosOneSigmaUpper\right]\deg$).
For this latter SHoES result we note that the face-off $\iota=180\deg$
orientation is just outside the 90\% confidence range.  It will be
particularly interesting to compare these constraints to those from
modeling of the short GRB, afterglow, and optical counterpart
associated with GW170817 \citep{GW170817MMA}.

We have presented a standard siren determination of the Hubble constant, using
a combination of a \ac{GW} distance and an \ac{EM} Hubble velocity estimate.
Our measurement does not use a ``distance ladder'', and makes no prior
assumptions about $H_0$.
We find
$H_0=\HnaughtMAPMax^{+\HnaughtMAPOneSigmaUpperDiff}_{-\HnaughtMAPOneSigmaLowerDiff}
\, \kms \, \Mpc^{-1}$, which is consistent with existing
measurements \citep{Riess2016,Planck2016}.
This first \ac{GW}--\ac{EM} multi-messenger event demonstrates the potential for
cosmological inference from \ac{GW} standard sirens.
We expect that additional multi-messenger binary neutron-star events will be detected in the coming years, and combining subsequent independent measurements of H0 from these future standard sirens will lead to an era of precision gravitational-wave cosmology.

\newpage

\section*{Methods}

\setcounter{figure}{0}
\renewcommand{\figurename}{Extended Data Figure}
\renewcommand{\tablename}{Extended Data Table}

\subsection*{Probability of optical counterpart association with NGC\,4993}
\label{sec:counterpartprob}

We calculate the probability that an NGC\,4993-like galaxy (or brighter) is
misidentified as the host by asking how often the centre of one or more such galaxies
falls by random chance within a given angular radius $\theta$ of the counterpart.
Assuming Poisson counting statistics this probability is given by
$P=1-\exp\left[-\pi \theta^2 S(<m)\right]$ where $S(<m)$ is the surface density
of galaxies with apparent magnitude equal to or brighter than $m$. From the
local galaxy sample distribution in the infrared (K-band) apparent magnitude \citep{Huang1998} we obtain
$S(<K)=0.68\times 10^{\left(0.64(K-10.0)-0.7\right)}\,\mathrm{deg}^{-2}$.
As suggested by \citep{Bloom2002}, we set $\theta$ equal to twice the half-light
radius of the galaxy, for which we use NGC\,4993's diameter of $\sim 1.1$
arcmin, as measured in the near infrared band (the predominant emission band
for early-type galaxies). Using $K=9.2$ mag taken from the 2MASS
survey \citep{2MASS2006} for NGC\,4993, we find
the probability of random chance association is $P=0.004$\%.

\subsection*{Finding the Hubble velocity of NGC\,4993}
\label{sec:hubblevelmethods}

In previous \ac{EM} determinations of the cosmic ``distance ladder",
the Hubble flow velocity of the local calibrating galaxies has generally been
estimated using redshift-independent secondary galaxy distance indicators, such
as the Tully-Fisher relation or type Ia supernovae, calibrated with more
distant samples that can be assumed to sit in quiet Hubble flow
 \citep{Freedman2001}.  We do not adopt this approach for NGC\,4993, however, in
order that our inference of the Hubble constant is fully independent of the
electromagnetic distance scale.  Instead we estimate the Hubble flow velocity
at the position of NGC\,4993 by correcting its measured recessional velocity
for local peculiar motions.

NGC\,4993 resides in a group of galaxies whose center-of-mass recession
velocity relative to the \ac{CMB} frame \citep{Hinshaw2009}
is \citep{Crook2007,Crook2008} $\recessionCluster \pm \recessionClusterUncert \,
\kms$.  We assume that all of the galaxies in the group are at the same
distance and therefore have the same Hubble flow velocity, which we assign to
be the Hubble velocity of GW170817. This assumption is accurate to within $1\%$
given that the radius of the group is $\sim 0.4$ Mpc. To calculate the Hubble
flow velocity of the group, we correct its measured recessional velocity by the
peculiar velocity caused by the local gravitational field. This is a
significant correction \citep{Springob2014,Carrick2015}; typical peculiar velocities are $300 \, \kms$, equivalent to
$\sim 10\%$ of the total recessional velocity at a distance of 40 Mpc.

We employ the 6dF galaxy redshift survey peculiar velocity
map \citep{Springob2014,Jones2009}, which used more than 8,000 Fundamental Plane
galaxies to map the peculiar velocity field in the Southern hemisphere out to
redshift $z \simeq 0.055$.  We weight the peculiar velocity corrections from this
catalog with a Gaussian kernel centered on NGC\,4993's sky position and with
a width of $8h^{-1}$\,\Mpc; the kernel width is independent of $H_0$
and is equivalent to a width of $800$\,\kms\ in velocity space, typical of the
widths used in the catalog itself. There are 10 galaxies in the 6dF peculiar
velocity catalog within one kernel width of NGC\,4993. In the \ac{CMB}
frame \citep{Hinshaw2009}, the weighted radial component of the peculiar velocity
and associated uncertainty is $\left\langle v_p \right\rangle = \pecvel \pm
\pecvelstderr \, \kms$.

We verified the robustness of this peculiar velocity correction by comparing it
with the velocity field reconstructed from the 2MASS redshift
survey \citep{Carrick2015,Huchra2012}.  This exploits the linear relationship
between the peculiar velocity and mass density fields smoothed on scales larger
than about $8 h^{-1}$ \, Mpc, and the constant of proportionality can be
determined by comparison with radial peculiar velocities of individual galaxies
estimated from e.g. Tully-Fisher and Type Ia supernovae distances.  Using these
reconstructed peculiar velocities, which have a larger associated
uncertainty \citep{Carrick2015} of $\hubblevelUncertcarrick \, \kms$, at the
position of NGC\,4993 we find a Hubble velocity in the \ac{CMB} frame of $v_H =
\hubblevelcarrick \,\kms$
-- in excellent agreement with the result derived using 6dF.
We adopt this larger uncertainty on the peculiar velocity correction in
recognition that the peculiar velocity estimated from the 6dF data may
represent an imperfect model of the true bulk flow at the location of
NGC\,4993.  For our inference of the Hubble constant we therefore use a Hubble
velocity $v_H = \HubbleVelMAPMax\pm\HubbleVelMAPOneSigmaUpperDiff\,\kms$ with
68.3\% uncertainty.

Finally, while we emphasise again the independence of our Hubble constant
inference from the electromagnetic distance scale, we note the consistency of
our GW distance estimate to NGC\,4993 with the Tully-Fisher distance estimate
derived by scaling back the Tully-Fisher relation calibrated with more distant
galaxies in quiet Hubble flow \citep{Sakai2000}. This also strongly supports the
robustness of our estimate for the Hubble velocity of NGC\,4993.

\subsection*{Summary of the model}
\label{sec:modelsumm}

Given observed data from a set of \ac{GW} detectors, $\gwdata$,
parameter estimation is used to generate a posterior on the parameters
that determine the waveform of the GW signal.  Parameters are inferred
within a Bayesian framework \citep{Veitch2015} by comparing strain
measurements \citep{GW170817} in the two LIGO detectors and the Virgo
detector with the gravitational waveforms expected from the inspiral
of two point masses \citep{Hannam2014} under general relativity.  We use
algorithms for removing short-lived detector noise
artifacts \citep{GW170817,Cornish2015} and we employ approximate
point-particle waveform
models \citep{Buonanno1998,Blanchet2014,Hannam2014}.  We have verified
that the systematic changes in the results presented here from
incorporating non-point-mass (tidal)
effects \citep{Hinderer2008,Vines2011} and from different data
processing methods are much smaller than the statistical uncertainties
in the measurement of $H_0$ and the binary orbital inclination angle.

From this analysis we can obtain the parameter estimation likelihood
of the observed \ac{GW} data, marginalized over all parameters
characterizing the GW signal except $d$ and $\cos\iota$,
\begin{align}
p(\gwdata \mid d,\cos\iota) = \int p(\gwdata \mid d,\cos\iota, \vec\lambda)
\,p(\vec\lambda) {\rm d}\vec\lambda.
\label{eq:GWlike}
\end{align}
The other waveform parameters are denoted by $\vec\lambda$, with
$p(\vec\lambda)$ denoting the corresponding prior.

Given perfect knowledge of the Hubble flow velocity of the \ac{GW} source, $v_H$, this posterior distribution can be readily converted into a posterior on $\cos\iota$ and $H_0 = v_H/d$,
\begin{multline}
p(H_0, \cos\iota | \gwdata) \\ \propto
(v_H/H_0^2) \, p(\gwdata \mid d = v_H/H_0,\cos\iota) \\ \times p_d(v_H/H_0)\,p_\iota(\cos\iota),
\end{multline}
where $p_d(d)$ and $p_\iota(\cos\iota)$ are the prior distributions on distance and inclination. For the Hubble velocity $v_H = \HubbleVelMAPMax\,\kms$, the maximum a posteriori distance from the \ac{GW} measurement of $\dgwPosteriorMAPMax \, \Mpc$ corresponds to $H_0 = 68.9 \,\kms \, \Mpc^{-1}$, so this procedure would be expected to generate a posterior on $H_0$ that peaks close to that value.

While the above analysis is conceptually straightforward, it makes a number of assumptions. In practice, the Hubble-flow velocity cannot be determined exactly and it must be corrected for uncertain peculiar velocities. The above does not explicitly set a prior on $H_0$, but instead inherits a $1/H_0^4$ prior from the usual $p_d(d) \propto d^2$ prior used in \ac{GW} parameter estimation. In addition, the logic in this model is that a redshift has been obtained first and the distance is then measured using \ac{GW}s. As \ac{GW} detectors cannot be pointed, we cannot target particular galaxies or redshifts for \ac{GW} sources. In practice, we wait for a \ac{GW} event to trigger the analysis and this introduces potential selection effects which we must consider. We will see below that the simple analysis described above does give results that are consistent with a more careful analysis for this first detection. However, the simple analysis cannot be readily extended to include second and subsequent detections, so we now describe a more general framework that does not suffer from these limitations.

\begin{figure}
\begin{centering}
  \includegraphics[width=\columnwidth]{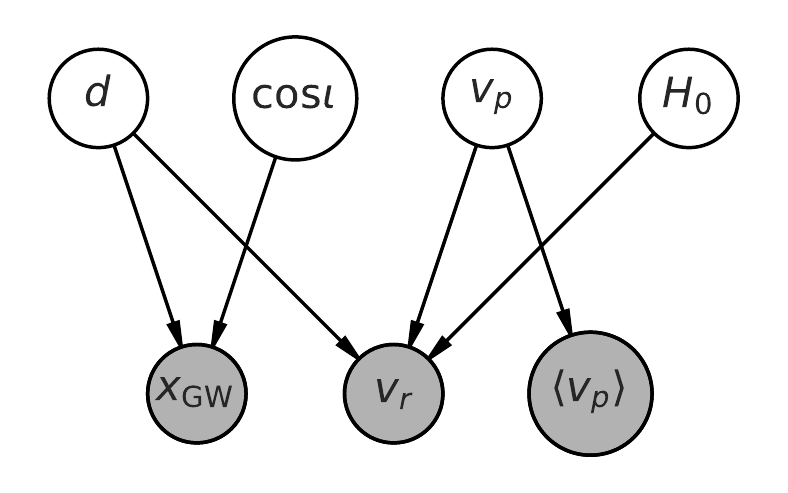}
\end{centering}
\caption{\label{fig:pgm} {\bf Graphical model illustrating the
    statistical relationships between the data and parameters.}  Open
  circles indicate parameters which require a prior; filled circles
  described measured data, which are conditioned on in the analysis.
  Here we assume we have measurements of the GW data, $\gwdata$, a
  recessional velocity (i.e.\ redshift), $v_r$, and the mean peculiar
  velocity in the neighborhood of NGC\,4993,
  $\left\langle v_p \right\rangle$.  Arrows flowing into a node
  indicate that the conditional probability density for the node
  depends on the source parameters; for example, the conditional
  distribution for the observed \ac{GW} data,
  $p\left( \gwdata \mid d, \cos \iota \right)$, discussed in the text,
  depends on the distance and inclination of the source (and
  additional parameters, here marginalized out).  }
\end{figure}

We suppose that we have observed a \ac{GW} event, which generated data $\gwdata$ in our detectors, and that we have also measured a recessional velocity for the host, $v_r$, and the peculiar velocity field, $\langle v_p \rangle$, in the vicinity of the host. These observations are statistically independent and so the combined likelihood is
\begin{multline}
p(\gwdata,v_r,\langle v_p\rangle \mid d, \cos\iota, v_p, H_0)=
\\ p(\gwdata \mid d,\cos\iota)\,p(v_r\mid d,v_p,H_0)\,p(\langle v_p\rangle\mid
v_p).
\label{eq:fulllike}
\end{multline}
The quantity $p(v_r\mid d,v_p,H_0)$ is the
likelihood of the recessional velocity measurement, which we model as
\begin{align}
\label{eq:recessional-conditional} p\left( v_r \mid d, v_p, H_0
\right) &= N\left[ v_p + H_0 d, \sigma^2_{v_r} \right]\left( v_r \right)
\end{align}
where $N\left[\mu, \sigma^2\right](x)$ is the normal (Gaussian) probability
density with mean $\mu$ and standard deviation $\sigma$ evaluated at $x$.  The
measured recessional velocity, $v_r = \recessionCluster \, \kms$, with
uncertainty $\sigma_{v_r} = \recessionClusterUncert \, \kms$, is the mean
velocity and standard error for the members of the group hosting NGC\,4993
taken from the \ac{2MASS} \citep{Crook2007,Crook2008}, corrected to the \ac{CMB}
frame \citep{Hinshaw2009}. We take a similar Gaussian likelihood for the measured
peculiar velocity, $\left\langle v_p \right\rangle = \pecvel \, \kms$, with
uncertainty $\sigma_{v_p} = \pecvelUncert \, \kms$:
\begin{align}
\label{eq:pecvel-conditional} p\left( \left\langle v_p
\right\rangle \mid v_p \right) = N\left[ v_p, \sigma^2_{v_p} \right]\left(
\left\langle v_p \right\rangle \right).
\end{align}
From the
likelihood~(\ref{eq:fulllike}) we derive the posterior
\begin{multline} p(H_0,d,\cos\iota,v_p\mid \gwdata ,v_r,\langle
v_p\rangle) \\ \propto \frac{p(H_0)}{{\cal N}_{\rm s}(H_0)}\,p(\gwdata \mid
d,\cos\iota)\,p(v_r\mid d,v_p,H_0) \\ \times\,p(\langle
v_p\rangle\mid v_p)\,p(d)\,p(v_p)\,p(\cos\iota), \label{eq:fullpost}
\end{multline}
where $p(H_0)$, $p(d)$, $p(v_p)$ and $p(\cos\iota)$ are the parameter prior
probabilities. Our standard analysis assumes a volumetric prior, $p\left( d \right)
\propto d^2$, on the Hubble distance, but we explore sensitivity to this
choice below. We take a flat-in-log prior on $H_0$, $p\left( H_0 \right)
\propto 1/H_0$, impose a flat (i.e.\,isotropic) prior on $\cos\iota$, and a
flat prior on $v_p$ for $v_p \in [-1000, 1000] \, \kms$. These priors characterise our beliefs about the
cosmological population of \ac{GW} events and their hosts before we make any additional measurements or account for selection biases. The full statistical model is summarized graphically
in Extended Data Figure~\ref{fig:pgm}. This model with these priors is our canonical analysis.

In Eq.~(\ref{eq:fullpost}), the term ${\cal N}_{\rm s}(H_0)$ encodes selection effects
 \citep{Loredo2004,Mandel2016,O1-BBH}. These arise
because of the finite sensitivity of our detectors. While all events in the Universe generate a response
in the detector, we will only be able to identify, and hence use, signals that generate a response of sufficiently high amplitude. The decision
about whether to include an event in the analysis is a property of the data only, in this case $\{\gwdata$,
$v_r$, $\langle v_p \rangle\}$, but the fact that we condition our analysis on a signal being detected, i.e.,
the data exceeding these thresholds, means that the likelihood must be renormalized to become the
likelihood for detected events. This is the role of 
\begin{multline}
{\cal N}_{\rm s}(H_0) = \int\limits_\mathrm{detectable}{\rm d}\vec\lambda\,{\rm d}d\,{\rm d}v_p\,{\rm d}\!\cos\iota\,{\rm d}\gwdata\,{\rm d}v_r\,{\rm d}\langle v_p\rangle \\  \times \left[ p(\gwdata \mid d,\cos\iota, \vec\lambda)\,p(v_r\mid d,v_p,H_0) \right. \\
 \left.\times\,p(\langle v_p\rangle\mid v_p)\,p(\vec\lambda)\,p(d)\,p(v_p)\,p(\cos\iota)\right]\,,
\label{eq:detfunc}
\end{multline}
where the integral is over the full prior ranges of the parameters, $\{d, v_p, \cos\iota, \vec\lambda\}$, and over data sets that would be selected for inclusion in the analysis, i.e., exceed the specified thresholds. If the integral was over all data sets it would evaluate to $1$, but because the range is restricted there can be a non-trivial dependence on parameters characterizing the population of sources, in this case $H_0$.

In the current analysis, there are in principle selection effects in both the
\ac{GW} data and the \ac{EM} data. However, around the time of detection of
GW170817, the LIGO-Virgo detector network had a detection horizon of
$\sim190\,$Mpc for \ac{BNS} events \citep{GW170817}, within which EM measurements
are largely complete. For example, the counterpart associated with GW170817 had
brightness $\sim17\,\mbox{mag}$ in the I band at
$40\,$Mpc \citep{DLT40,LCO,VISTA,MASTER,1M2H}; this
source would be $\sim22\,\mbox{mag}$ at $400\,$Mpc, and thus still detectable
by survey telescopes such as DECam well beyond the \ac{GW} horizon. Even the dimmest
theoretical lightcurves for kilonovae are expected to peak at
$\sim22.5\,\mbox{mag}$ at the LIGO--Virgo horizon \citep{2012ApJ...746...48M}. We
therefore expect that we are dominated by \ac{GW} selection effects at the current
time and can ignore \ac{EM} selection effects. The fact that the fraction of
\ac{BNS} events that will have observed kilonova counterparts is presently unknown
does not modify these conclusions, since we can restrict our analysis to
\ac{GW} events with kilonova counterparts only.

In the \ac{GW} data, the decision about whether or not to analyse an
event is largely determined by the signal-to-noise ratio (SNR),
$\rho$, of the event. A reasonable model for the selection process is a
cut in SNR, i.e., events with $\rho > \rho_*$ are analysed \citep{GW150914-Rate-Supp}. In that model, the integral over $\gwdata$ in Eq.~(\ref{eq:detfunc}) can be replaced by an integral over SNR from $\rho_*$ to $\infty$, and
$p(\gwdata | d,\cos\iota,\vec\lambda)$ replaced by
$p(\rho | d,\cos\iota,\vec\lambda)$ in the
integrand.
This distribution depends on the noise properties of the operating detectors,
and on the intrinsic strain amplitude of the source. The former are clearly
independent of the population parameters, while the latter scales like a
function of the source parameters divided by the luminosity distance. The
dependence on source parameters is on redshifted parameters, which introduces
an explicit redshift dependence. However, within the $\sim190\,$Mpc horizon,
redshift corrections are at most $\lesssim 5\%$, and the Hubble constant
measurement is a weak function of these, meaning the overall impact is even
smaller. At present, whether or not a particular event in the population ends
up being analysed can therefore be regarded as a function of $d$ only. When
\ac{GW} selection effects dominate, only the terms in Eq.~(\ref{eq:detfunc})
arising from the \ac{GW} measurement matter. As these are a function of $d$
only and we set a prior on $d$, there is no explicit $H_0$ dependence in these
terms. Hence, ${\cal N}_{\rm s}(H_0)$ is a constant and can be ignored. This
would not be the case if we set a prior on the redshifts of potential sources
instead of their distances, since then changes in $H_0$ would modify the range
of detectable redshifts. As the LIGO--Virgo detectors improve in sensitivity
the redshift dependence in the \ac{GW} selection effects will become more
important, as will EM selection effects. However, at that point we will also
have to consider deviations in the cosmological model from the simple Hubble
flow described in Eq.~(\ref{eq:hubble}) of the main article.

Marginalising Eq.~(\ref{eq:fullpost}) over $d$, $v_p$ and $\cos\iota$ then yields
\begin{multline}
\label{eq:masterh0inference}
p(H_0\mid \gwdata,v_r,\langle v_p\rangle) \propto p(H_0)\int {\rm d}d\,{\rm d}v_p\,{\rm d}\!\cos\iota \\ \times p(\gwdata \mid d,\cos\iota)\,p(v_r\mid d,v_p,H_0) \\ \times p(\langle v_p\rangle\mid v_p)\,p(d)\,p(v_p)\,p(\cos\iota)\,.
\end{multline}
The posterior computed in this way was shown in Figure~\ref{fig:killer-plot1} in the main article and has a maximum a posteriori value and minimal $68.3\%$ credible interval of $\HnaughtMAPMax^{+\HnaughtMAPOneSigmaUpperDiff}_{-\HnaughtMAPOneSigmaLowerDiff}  \, \kms \, \Mpc^{-1}$, as quoted in the main article. The posterior mean is $\Hnaught\, \kms \, \Mpc^{-1}$ and the standard deviation is $\HnaughtUncert\, \kms \, \Mpc^{-1}$. Various other summary statistics are given in Extended Data Table~\ref{tab:results}.

\subsection*{Robustness to prior specification}
\label{sec:robustness}

Our canonical analysis uses a uniform volumetric prior on distance, $p(d) \propto
d^2$. The distribution of galaxies is not completely uniform due to
clustering, so we explore sensitivity to this prior choice. We are
free to place priors on any two of the three variables
$\{d, H_0, z\}$, where $z = H_0 d/c$ is the Hubble flow redshift of
NGC\,4993. A choice of prior for two of these variables induces a prior
on the third which may or may not correspond to a natural choice for
that parameter. A prior on $z$ could be obtained from galaxy catalog
observations \citep{Dalya20016}, but must be corrected for incompleteness. When setting a prior on $H_0$ and $z$, the
posterior becomes
\begin{multline}
p(H_0,z,\cos\iota,v_p\mid \gwdata,v_r,\langle v_p\rangle) \\ \propto \frac{p(H_0)}{{\cal N}_{\rm s}(H_0)}\, p(\gwdata\mid d=cz/H_0,\cos\iota)\,p(v_r\mid z,v_p) \\ \times\,p(\langle v_p\rangle\mid v_p)\,p(z)\,p(v_p)\,p(\cos\iota),
\end{multline}
but now
\begin{multline}
{\cal N}_{\rm s}(H_0) =  \int\limits_\mathrm{detectable} {\rm d}z\,{\rm d}v_p\,{\rm d}\!\cos\iota\,{\rm d}x_{GW}\,{\rm d}v_r\,{\rm d}\langle v_p\rangle \\ \times p(\gwdata\mid d=cz/H_0,\cos\iota)\,p(v_r\mid z,v_p)  \\
\times\,p(\langle v_p\rangle\mid v_p)\,p(z)\,p(v_p)\,p(\cos\iota)\,.
\end{multline}
When \ac{GW} selection effects dominate, the integral is effectively
\begin{multline}
{\cal N}_{\rm s}(H_0) = \int {\rm d}z\,{\rm d}\!\cos\iota\,{\rm d}x_{GW} \\ \times p(\gwdata\mid d=cz/H_0,\cos\iota) p(z)\,p(\cos\iota)\, \\
=  \int {\rm d}d\,{\rm d}\!\cos\iota\,{\rm d}x_{GW} \\ \times p(\gwdata\mid d,\cos\iota)p(d H_0/c)\,p(\cos\iota)\,(H_0/c)\,,
\end{multline}
which has an $H_0$ dependence, unless $p(z)$ takes a special,
$H_0$-dependent form, $p(z) = f(z/H_0)/H_0$. However, if
the redshift prior is volumetric, $p(z) \propto z^2$, the selection
effect term is $\propto H_0^3$, which cancels a similar correction to
the likelihood and gives a posterior on $H_0$ that is identical to the
canonical analysis.

\begin{figure}
  \begin{centering}
  \includegraphics[width=\columnwidth]{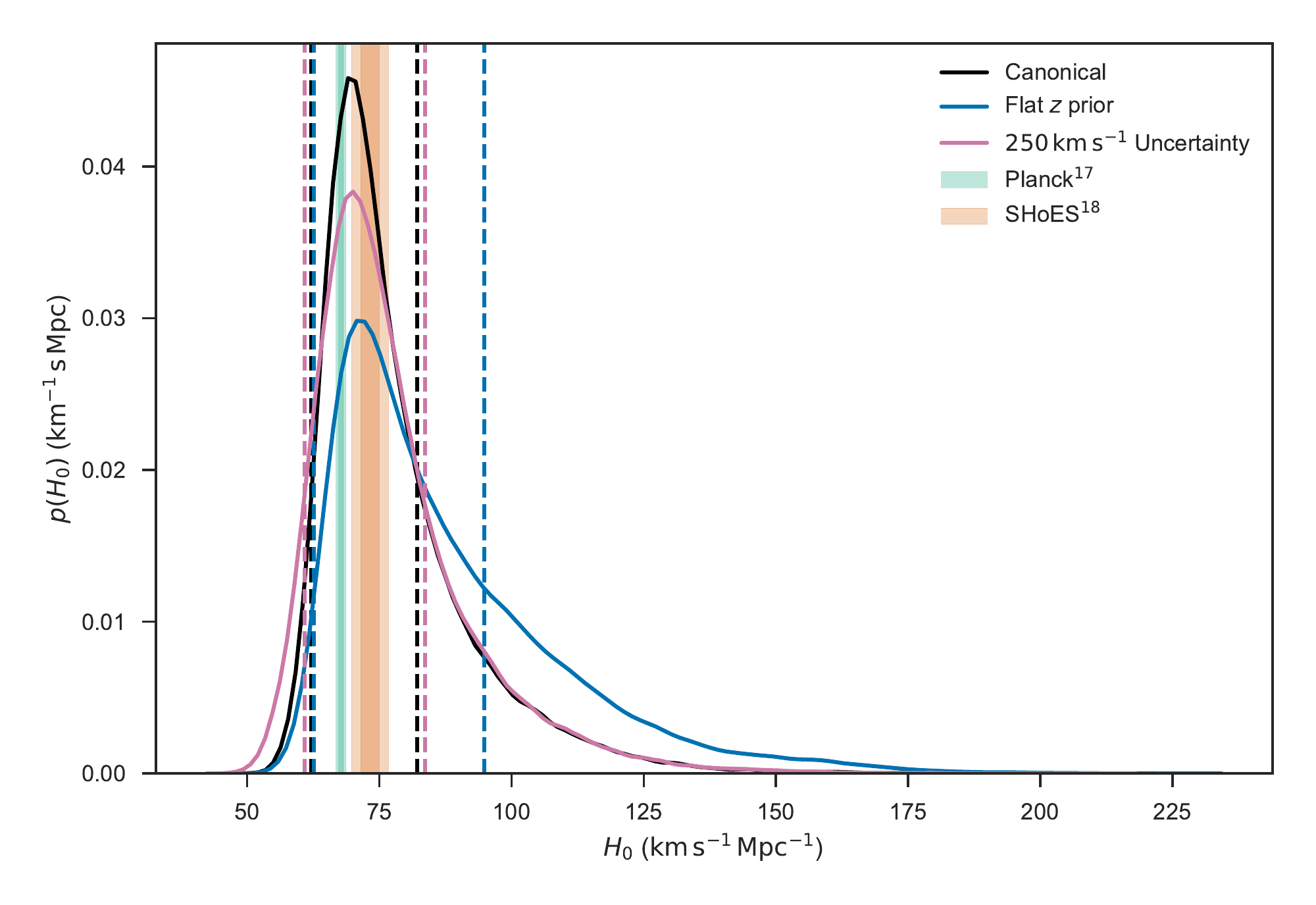}
  \end{centering}
  \caption{\label{fig:prior-sensitivity} \textbf{Using different
      assumptions compared to our canonical analysis.}  The posterior
    distribution on $H_0$ discussed in the main text is shown in
    black, the alternative flat prior on $z$ (discussed in the Methods
    section) gives the distribution shown in blue, and the increased
    uncertainty ($250 \, \mathrm{km} \, \mathrm{s}^{-1}$) applied to
    our peculiar velocity measurement (also discussed in the Methods
    section) is shown in pink.  Minimal $68.3\%$ ($1\sigma$) credible
    intervals are shown by dashed lines.  }
\end{figure}

For a single event, any choice of prior can be mapped to our canonical
analysis with a different prior on $H_0$. For any reasonable prior
choices on $d$ or $z$, we would expect to gradually lose sensitivity
to the particular prior choice as further observed events are added to
the analysis. However, to illustrate the uncertainty that comes from
the prior choice for this first event, we compare in Extended Data
Figure~\ref{fig:prior-sensitivity} and Extended Data
Table~\ref{tab:results} the results from the canonical prior choice
$p\left( d \right) \propto d^2$ to those from two other choices: using
a flat prior on $z$, and assuming a velocity correction due to the
peculiar velocity of NGC\,4993 that is a Gaussian with width
$250 \, \mathrm{km} \, \mathrm{s}^{-1}$.  (To do the first of these,
the posterior samples from \ac{GW} parameter estimation have to be
re-weighted, since they are generated with the $d^2$ prior used in the
canonical analysis. We first ``undo'' the default prior before
applying the desired new prior.)

The choice of a flat prior on $z$ is motivated by the simple model
described above, in which we imagine first making a redshift
measurement for the host and then use that as a prior for analysing
the \ac{GW} data. Setting priors on distance and redshift, the simple
analysis gives the same result as the canonical analysis, but now we
set a prior on redshift and $H_0$ and obtain a different result. This
is to be expected because we are making different assumptions about
the underlying population, and it arises for similar reasons as the
different biases in peculiar velocity measurements based on
redshift-selected or distance-selected
samples \citep{1995PhR...261..271S}. As can be seen in Extended Data
Table~\ref{tab:results}, the results change by less than $1\sigma$, as
measured by the statistical error of the canonical analysis.

By increasing the uncertainty in the peculiar velocity prior, we test the assumptions
in our canonical analysis that (1) NGC\,4993 is a member of
the nearby group of galaxies, and (2) that this group
has a center-of-mass velocity close to the Hubble flow. The
results in Extended Data Table~\ref{tab:results} summarizes
changes in the values of $H_0$ and in the error bars.

We conclude that the impact of a reasonable change to the prior is small
relative to the statistical uncertainties for this event.

\begin{deluxetable*}{lrrrr}

\tablecolumns{5}

\tablecaption{\label{tab:results} Summary of constraints on the Hubble
constant, binary inclination, and distance}

\tablehead{\colhead{Parameter} & \colhead{68.3\% Symm.} & \colhead{68.3\% MAP} &
\colhead{90\% Symm.} & \colhead{90\% MAP}}
\startdata
$H_0 / \left( \mathrm{km} \, \mathrm{s}^{-1} \, \mathrm{Mpc}^{-1} \right)$ & $\HnaughtMedian^{+\HnaughtOneSigmaUpperDiff}_{-\HnaughtOneSigmaLowerDiff}$ & $\HnaughtMAPMax^{+\HnaughtMAPOneSigmaUpperDiff}_{-\HnaughtMAPOneSigmaLowerDiff}$ & $\HnaughtMedian^{+\HnaughtNinetyUpperDiff}_{-\HnaughtNinetyLowerDiff}$ & $\HnaughtMAPMax^{+\HnaughtMAPNinetyUpperDiff}_{-\HnaughtMAPNinetyLowerDiff}$\\
$H_0 / \left( \mathrm{km} \, \mathrm{s}^{-1} \, \mathrm{Mpc}^{-1} \right)$ (flat in $z$ prior) & $\HnaughtFlatZMedian^{+\HnaughtFlatZOneSigmaUpperDiff}_{-\HnaughtFlatZOneSigmaLowerDiff}$ & $\HnaughtFlatZMAPMax^{+\HnaughtFlatZMAPOneSigmaUpperDiff}_{-\HnaughtFlatZMAPOneSigmaLowerDiff}$ & $\HnaughtFlatZMedian^{+\HnaughtFlatZNinetyUpperDiff}_{-\HnaughtFlatZNinetyLowerDiff}$ & $\HnaughtFlatZMAPMax^{+\HnaughtFlatZMAPNinetyUpperDiff}_{-\HnaughtFlatZMAPNinetyLowerDiff}$ \\
$H_0 / \left( \mathrm{km} \, \mathrm{s}^{-1} \, \mathrm{Mpc}^{-1} \right)$ ($250 \, \mathrm{km} \, \mathrm{s}^{-1}$ $\sigma_{v_r}$) & $\HnaughtLargeVelocityMedian^{+\HnaughtLargeVelocityOneSigmaUpperDiff}_{-\HnaughtLargeVelocityOneSigmaLowerDiff}$ & $\HnaughtLargeVelocityMAPMax^{+\HnaughtLargeVelocityMAPOneSigmaUpperDiff}_{-\HnaughtLargeVelocityMAPOneSigmaLowerDiff}$ & $\HnaughtLargeVelocityMedian^{+\HnaughtLargeVelocityNinetyUpperDiff}_{-\HnaughtLargeVelocityNinetyLowerDiff}$ & $\HnaughtLargeVelocityMAPMax^{+\HnaughtLargeVelocityMAPNinetyUpperDiff}_{-\HnaughtLargeVelocityMAPNinetyLowerDiff}$\\
$\cos\iota$ (GW only) & $\cosiotaMedian^{+\cosiotaOneSigmaUpperDiff}_{-\cosiotaOneSigmaLowerDiff}$ & $\cosiotaMAPMax^{+\cosiotaMAPOneSigmaUpperDiff}_{-\cosiotaMAPOneSigmaLowerDiff}$ & $\cosiotaMedian^{+\cosiotaNinetyUpperDiff}_{-\cosiotaNinetyLowerDiff}$ & $\cosiotaMAPMax^{+\cosiotaMAPNinetyUpperDiff}_{-\cosiotaMAPNinetyLowerDiff}$\\
$\cos\iota$ (SHoES) & $\riessCosIotaMedian^{+\riessCosIotaOneSigmaUpperDiff}_{-\riessCosIotaOneSigmaLowerDiff}$ & $\riessCosIotaMAPMax^{+\riessCosIotaMAPOneSigmaUpperDiff}_{-\riessCosIotaMAPOneSigmaLowerDiff}$ & $\riessCosIotaMedian^{+\riessCosIotaNinetyUpperDiff}_{-\riessCosIotaNinetyLowerDiff}$ & $\riessCosIotaMAPMax^{+\riessCosIotaMAPNinetyUpperDiff}_{-\riessCosIotaMAPNinetyLowerDiff}$ \\
$\cos\iota$ (Planck) &
$\planckCosIotaMedian^{+\planckCosIotaOneSigmaUpperDiff}_{-\planckCosIotaOneSigmaLowerDiff}$
&
$\planckCosIotaMAPMax^{+\planckCosIotaMAPOneSigmaUpperDiff}_{-\planckCosIotaMAPOneSigmaLowerDiff}$
&
$\planckCosIotaMedian^{+\planckCosIotaNinetyUpperDiff}_{-\planckCosIotaNinetyLowerDiff}$
&
$\planckCosIotaMAPMax^{+\planckCosIotaMAPNinetyUpperDiff}_{-\planckCosIotaMAPNinetyLowerDiff}$\\
$\iota/\mathrm{deg}$ (GW only) & $\cosiotaArcCosMedian^{+\cosiotaArcCosOneSigmaUpperDiff}_{-\cosiotaArcCosOneSigmaLowerDiff}$ & $\cosiotaMAPArcCosMax^{+\cosiotaMAPArcCosOneSigmaUpperDiff}_{-\cosiotaMAPArcCosOneSigmaLowerDiff}$ & $\cosiotaArcCosMedian^{+\cosiotaArcCosNinetyUpperDiff}_{-\cosiotaArcCosNinetyLowerDiff}$ & $\cosiotaMAPArcCosMax^{+\cosiotaMAPArcCosNinetyUpperDiff}_{-\cosiotaMAPArcCosNinetyLowerDiff}$\\
$\iota/\mathrm{deg}$ (SHoES) & $\riessCosIotaArcCosMedian^{+\riessCosIotaArcCosOneSigmaUpperDiff}_{-\riessCosIotaArcCosOneSigmaLowerDiff}$ & $\riessCosIotaMAPArcCosMax^{+\riessCosIotaMAPArcCosOneSigmaUpperDiff}_{-\riessCosIotaMAPArcCosOneSigmaLowerDiff}$ & $\riessCosIotaArcCosMedian^{+\riessCosIotaArcCosNinetyUpperDiff}_{-\riessCosIotaArcCosNinetyLowerDiff}$ & $\riessCosIotaMAPArcCosMax^{+\riessCosIotaMAPArcCosNinetyUpperDiff}_{-\riessCosIotaMAPArcCosNinetyLowerDiff}$ \\
$\iota/\mathrm{deg}$ (Planck) &
$\planckCosIotaArcCosMedian^{+\planckCosIotaArcCosOneSigmaUpperDiff}_{-\planckCosIotaArcCosOneSigmaLowerDiff}$
&
$\planckCosIotaMAPArcCosMax^{+\planckCosIotaMAPArcCosOneSigmaUpperDiff}_{-\planckCosIotaMAPArcCosOneSigmaLowerDiff}$
&
$\planckCosIotaArcCosMedian^{+\planckCosIotaArcCosNinetyUpperDiff}_{-\planckCosIotaArcCosNinetyLowerDiff}$
&
$\planckCosIotaMAPArcCosMax^{+\planckCosIotaMAPArcCosNinetyUpperDiff}_{-\planckCosIotaMAPArcCosNinetyLowerDiff}$\\
$d / \left( \mathrm{Mpc} \right)$ &
$\dgwPosteriorMedian^{+\dgwPosteriorOneSigmaUpperDiff}_{-\dgwPosteriorOneSigmaLowerDiff}$ &
$\dgwPosteriorMAPMax^{+\dgwPosteriorMAPOneSigmaUpperDiff}_{
-\dgwPosteriorMAPOneSigmaLowerDiff}$ &
$\dgwPosteriorMedian^{+\dgwPosteriorNinetyUpperDiff}_{-\dgwPosteriorNinetyLowerDiff}$ &
$\dgwPosteriorMAPMax^{+\dgwPosteriorNinetyUpperDiff}_{-\dgwPosteriorMAPNinetyLowerDiff}$\\
\enddata
\tablecomments{We give both one-sigma (68.3\%)
  and 90\% credible intervals for each quantity.  ``Symm.'' refers to
  a symmetric interval (e.g.\ median and 5\% to 95\% range), while
  ``MAP'' refers to maximum a posteriori intervals (e.g.\ MAP value
  and smallest range enclosing 90\% of the posterior).  Values given
  for $\iota$ are derived from arc-cosine transforming the
  corresponding values for $\cos \iota$, so the ``MAP'' values differ
  from those that would be derived from the posterior on $\iota$.}
\end{deluxetable*}

\subsection*{Incorporating additional constraints on $H_0$}
\label{sec:addconst}

By including previous measurements of $H_0$ \citep{Planck2016,Riess2016} we can constrain the orbital inclination more precisely. We do this by setting the $H_0$ prior in Eq.~(\ref{eq:fullpost}) to $p(H_0 | \mu_{H_0}, \sigma^2_{H_0}) = N[\mu_{H_0}, \sigma^2_{H_0}]$, where for ShoES \citep{Riess2016} $\mu_{H_0} = \riessHnaught\, \kms \, \Mpc^{-1}$ and $\sigma_{H_0} =  \riessHnaughtUncert \, \kms \, \Mpc^{-1}$, while for Planck \citep{Planck2016} $\mu_{H_0}= \planckHnaught \, \kms \, \Mpc^{-1}$ and $ \sigma_{H_0} =  \planckHnaughtUncert \, \kms \, \Mpc^{-1}$. The posterior on $\cos\iota$ is then
\begin{multline}
\label{eq:cosiposterior}
p(\cos\iota\mid \gwdata,v_r,\langle v_p\rangle, \mu_{H_0},\sigma^2_{H_0}) \propto \int {\rm d}d\,{\rm d}v_p\,{\rm d}\!H_0 \\ \times p(\gwdata \mid d,\cos\iota)\,p(v_r\mid d,v_p,H_0)\,p(\langle v_p\rangle\mid v_p) \\ \times\,p(H_0 | \mu_{H_0}, \sigma^2_{H_0})\,p(d)\,p(v_p)\,.
\end{multline}
This posterior was shown in Figure~\ref{fig:inclination} of the main article.

\acknowledgments

The authors gratefully acknowledge the support of the United States
National Science Foundation (NSF) for the construction and operation of the
LIGO Laboratory and Advanced LIGO as well as the Science and Technology Facilities Council (STFC) of the
United Kingdom, the Max-Planck-Society (MPS), and the State of
Niedersachsen/Germany for support of the construction of Advanced LIGO 
and construction and operation of the GEO600 detector. 
Additional support for Advanced LIGO was provided by the Australian Research Council.
The authors gratefully acknowledge the Italian Istituto Nazionale di Fisica Nucleare (INFN),  
the French Centre National de la Recherche Scientifique (CNRS) and
the Foundation for Fundamental Research on Matter supported by the Netherlands Organisation for Scientific Research, 
for the construction and operation of the Virgo detector
and the creation and support  of the EGO consortium. 
The authors also gratefully acknowledge research support from these agencies as well as by 
the Council of Scientific and Industrial Research of India, 
the Department of Science and Technology, India,
the Science \& Engineering Research Board (SERB), India,
the Ministry of Human Resource Development, India,
the Spanish  Agencia Estatal de Investigaci\'on,
the  Vicepresid\`encia i Conselleria d'Innovaci\'o, Recerca i Turisme and the Conselleria d'Educaci\'o i Universitat del Govern de les Illes Balears,
the Conselleria d'Educaci\'o, Investigaci\'o, Cultura i Esport de la Generalitat Valenciana,
the National Science Centre of Poland,
the Swiss National Science Foundation (SNSF),
the Russian Foundation for Basic Research, 
the Russian Science Foundation,
the European Commission,
the European Regional Development Funds (ERDF),
the Royal Society, 
the Scottish Funding Council, 
the Scottish Universities Physics Alliance, 
the Hungarian Scientific Research Fund (OTKA),
the Lyon Institute of Origins (LIO),
the National Research, Development and Innovation Office Hungary (NKFI), 
the National Research Foundation of Korea,
Industry Canada and the Province of Ontario through the Ministry of Economic Development and Innovation, 
the Natural Science and Engineering Research Council Canada,
the Canadian Institute for Advanced Research,
the Brazilian Ministry of Science, Technology, Innovations, and Communications,
the International Center for Theoretical Physics South American Institute for Fundamental Research (ICTP-SAIFR), 
the Research Grants Council of Hong Kong,
the National Natural Science Foundation of China (NSFC),
the Leverhulme Trust, 
the Research Corporation, 
the Ministry of Science and Technology (MOST), Taiwan
and
the Kavli Foundation.
The authors gratefully acknowledge the support of the NSF, STFC, MPS, INFN, CNRS and the
State of Niedersachsen/Germany for provision of computational resources.
 This article has been assigned the document number
\href{http://dcc.ligo.org/P1700296/public}{LIGO-P1700296}.

We thank the University of Copenhagen, DARK Cosmology Centre, and the Niels Bohr International Academy for hosting D.A.C., R.J.F., A.M.B., E.R., and M.R.S.\ during the discovery of GW170817/SSS17a.  R.J.F., A.M.B., E.R., and D.E.H.\ were participating in the Kavli Summer Program in Astrophysics, ``Astrophysics with gravitational wave detections.''  This program was supported by the the Kavli Foundation, Danish National Research Foundation, the Niels Bohr International Academy, and the DARK Cosmology Centre.

The UCSC group is supported in part by NSF grant AST--1518052, the Gordon \& Betty Moore Foundation, the Heising-Simons Foundation, generous donations from many individuals through a UCSC Giving Day grant, and from fellowships from the Alfred P.\ Sloan Foundation (R.J.F), the David and Lucile Packard Foundation (R.J.F.\ and E.R.) and the Niels Bohr Professorship from the DNRF (E.R.).
A.M.B.\ acknowledges support from a UCMEXUS-CONACYT Doctoral Fellowship.
Support for this work was provided by NASA through Hubble Fellowship
grants HST--HF--51348.001 and HST--HF--51373.001 awarded by the Space Telescope Science Institute, which is operated by the Association of Universities for Research in Astronomy, Inc., for NASA, under contract NAS5--26555.

The Berger Time-Domain Group at Harvard is supported in part by the NSF through grants AST-1411763 and AST-1714498, and by NASA through grants NNX15AE50G and NNX16AC22G.

Funding for the DES Projects has been provided by the DOE and NSF (USA), MEC, MICINN, MINECO (Spain), STFC (UK), HEFCE (UK), NCSA (UIUC), KICP (U. Chicago), CCAPP (Ohio State),
MIFPA (Texas A\&M), CNPQ, FAPERJ, FINEP (Brazil), DFG (Germany) and the Collaborating Institutions in the Dark Energy Survey.
The Collaborating Institutions are Argonne Lab, UC Santa Cruz, University of Cambridge, CIEMAT-Madrid, University of Chicago, University College London,
DES-Brazil Consortium, University of Edinburgh, ETH Z{\"u}rich, Fermilab, University of Illinois, ICE (IEEC-CSIC), IFAE Barcelona, Lawrence Berkeley Lab,
LMU M{\"u}nchen and the associated Excellence Cluster Universe, University of Michigan, NOAO, University of Nottingham, Ohio State University, University of
Pennsylvania, University of Portsmouth, SLAC National Lab, Stanford University, University of Sussex, Texas A\&M University, and the OzDES Membership Consortium.
Based in part on observations at Cerro Tololo Inter-American Observatory, National Optical Astronomy Observatory, which is operated by the Association of
Universities for Research in Astronomy (AURA) under a cooperative agreement with the National Science Foundation.
The DES Data Management System is supported by the NSF under Grant Numbers AST-1138766 and AST-1536171. The DES participants from Spanish institutions are partially
supported by MINECO under grants AYA2015-71825, ESP2015-88861, FPA2015-68048, and Centro de Excelencia SEV-2012-0234, SEV-2016-0597 and MDM-2015-0509. Research leading
to these results has received funding from the ERC under the EU's 7$^{\rm th}$ Framework Programme including grants ERC 240672, 291329 and 306478.
We acknowledge support from the Australian Research Council Centre of Excellence for All-sky Astrophysics (CAASTRO), through project number CE110001020.
This manuscript has been authored by Fermi Research Alliance, LLC under Contract No. DE-AC02-07CH11359 with the U.S. Department of Energy, Office of Science, Office of High Energy Physics. The United States Government retains and the publisher, by accepting the article for publication, acknowledges that the United States Government retains a non-exclusive, paid-up, irrevocable, world-wide license to publish or reproduce the published form of this manuscript, or allow others to do so, for United States Government purposes.

D.J.S. acknowledges support for the DLT40 program from NSF grant  AST-1517649.

Support for I.A. was provided by NASA through the Einstein Fellowship Program, grant PF6-170148. G.H., D.A.H. and C.M. are supported by NSF grant AST-1313484. D.P. acknowledges support by Israel Science Foundation grant 541/17.

VINROUGE is an European Southern Observatory Large Survey (id: 0198.D-2010).

MASTER acknowledges the Lomonosov MSU Development Programm and the Russian Federation Ministry of Education and Science.

   This research has made use of the NASA/IPAC Extragalactic Database
   (NED) which is operated by the Jet Propulsion Laboratory,
   California Institute of Technology, under contract with the
   National Aeronautics and Space Administration.

All authors contributed to the
   work presented in this paper.

The authors declare that they have no
competing financial interests.

Correspondence and requests for materials should
  be addressed to the LVC spokespeople~(email:
  \href{mailto:lsc-spokesperson@ligo.org}{lsc-spokesperson@ligo.org},
  \href{mailto:virgo-spokesperson@ego-gw.eu}{virgo-spokesperson@ego-gw.eu}).

Available public codes can be found at the
  LIGO Open Science Center (\url{https://losc.ligo.org}).

Available public data can be found at the
  LIGO Open Science Center (\url{https://losc.ligo.org}).

\bibliography{masterbib,gcn}

\allauthors

\end{document}